\documentclass[authorversion, screen, nonacm]{acmart}

\usepackage{colortbl}
\usepackage{xcolor}
\usepackage{multirow}
\usepackage{ulem} 
\usepackage{subfigure}

\AtBeginDocument{%
  \providecommand\BibTeX{{%
    \normalfont B\kzern-0.5em{\scshape i\kern-0.25em b}\kern-0.8em\TeX}}}

\begin{document}

\title{Chatbots language design: the influence of language variation on user experience}

\author{Ana Paula Chaves}
\email{anachaves@utfpr.edu.br}
\orcid{0000-0002-2307-3099}
\affiliation{%
  \institution{Northern Arizona University}
  \streetaddress{1295 S Knoles Dr}
  \city{Flagstaff}
  \state{Arizona}
  \postcode{86011}
}
\affiliation{%
  \institution{Fed. University of Technology--Paran\'a}
  \streetaddress{R. Rosalina Maria dos Santos, 1233}
  \city{Campo Mour\~ao}
  \country{Brazil}
}

\author{Jesse Egbert}
\email{Jesse.Egbert@nau.edu}
\affiliation{%
  \institution{Department of English, Northern Arizona University}
  \streetaddress{705 S Beaver St}
  \city{Flagstaff}
  \state{Arizona}
  \postcode{86011}
}
\author{Toby Hocking}
\email{Toby.Hocking@nau.edu}
\author{Eck Doerry}
\email{Eck.Doerry@nau.edu}
\author{Marco Aurelio Gerosa}
\email{Marco.Gerosa@nau.edu}
\affiliation{%
  \institution{School of Informatics, Computing, and Cyber Systems, Northern Arizona University}
}

\renewcommand{\shortauthors}{Chaves, et al.}

\begin{abstract}
Chatbots are often designed to mimic social roles attributed to humans. However, little is known about the impact on user’s perceptions of using language that fails to conform to the associated social role. Our research draws on sociolinguistic theory to investigate how a chatbot’s language choices can adhere to the expected social role the agent performs within a given context. In doing so, we seek to understand whether chatbots design should account for linguistic register. This research analyzes how register differences play a role in shaping the user's perception of the human-chatbot interaction. Ultimately, we want to determine whether register-specific language influences users' perceptions and experiences with chatbots. We produced parallel corpora of conversations in the tourism domain with similar content and varying register characteristics and evaluated users' preferences of chatbot's linguistic choices in terms of appropriateness, credibility, and user experience. Our results show that register characteristics are strong predictors of user's preferences, which points to the needs of designing chatbots with register-appropriate language to improve acceptance and users' perceptions of chatbot interactions. 
\end{abstract}



\maketitle

\section{Introduction}
\label{sec:introduction}

Recent advances in conversational technologies have promoted the increasing popularity of chatbots~\cite{walgama2017chatbots}, which are disembodied conversational interfaces that interact with users in natural language via a text-based messaging interface. A recent report on the chatbot market~\cite{grandviewresearch2017chatbot}  attests to their increasing demand, predicting a global chatbot market of USD 1.25 billion by 2025. The skyrocketing interest in chatbot technologies has brought new challenges for the HCI field~\cite{brandtzaeg2018chatbots, folstad2017chatbots, neururer2018perceptions} and, despite improvements in design, users may not always be satisfied with their experiences~\cite{komatsu2012does, luger2016like}, which may affect their attitudes towards the technology~\cite{kiseleva2016predicting}.

For chatbots, natural language conversation is the primary mechanism for achieving interactional goals. Therefore, developing a more comprehensive understanding of the linguistic design of chatbot conversations and its effects on users' perceptions is critical to the success of chatbot technologies. Previous research on chatbot design suggests that when chatbots misuse language (e.g., conveying excessive (in)formality or using incoherent style), the conversation sounds strange to the user, and leads to frustration~\cite{duijst2017can, kirakowski2009establishing, mairesse2009can}. To date, language design for chatbots has focused primarily on ensuring that chatbots produce coherent and grammatically correct responses, and on improving functional performance and accuracy (see e.g.~\cite{jiang2017towards, maslowski2017wild, massaro1999developing, zdravkova2000conceptual}). Although current chatbots may, at some functional level, provide users with the answers they seek, the utterances portray arbitrary patterns of language that may not take into account the interactional situation. For instance, one would expect a chatbot representing a financial advisor to employ a different tone and linguistic features than that employed by a chatbot advising teens on current fashion choices. Currently, the design of a particular chatbot's linguistic choices is often based on ad-hoc analyses of user characteristics or the chatbot's persona. For machine language generation, models are trained using available corpora in the target domain, but they do not consider the particular context of the corpora's conversations. 

Little is known about the effect of these design decisions on users' perceptions, much less about how to tailor chatbot design to the particular situation of use. When exploring a list of key factors that could influence a user's perceptions of chatbots, scholars even argued that using an appropriate language style is not relevant as long as the user can understand the chatbot's answer~\citep{balaji2019assessing, bocker2019usability}. In contrast, empirical studies have repeatedly demonstrated that chatbot's linguistic choices influence users' perceptions and behavior toward chatbots~\citep{elsholz2019exploring, araujo2018living, tariverdiyeva2019chatbots, resendez2020very, thomas2018style}. Using appropriate linguistic choices potentially increases human-likeness~\cite{hill2015real, jenkins2007analysis, gnewuch2017towards} and believability~\cite{jenkins2007analysis, morris2002conversational, morrissey2013realness, tallyn2018ethnobot}, as well as enhance the overall perception of the quality of the interaction~\cite{jakic2017impact}. ~\citet{resendez2020very} showed that a chatbot's linguistic style evoke competence and trust as well as likeability and usefulness. Developing a strong basis for designing not just what a chatbot says but also how it says it must be a priority for creating the next generation of chatbots. This research establishes a framework for analyzing the effect of linguistic choices on users' perceptions, and takes a first step toward developing a prescriptive basis for tailoring chatbot linguistic choices to specific interactional situations.

Humans have developed a sense of how to adapt their tone, idioms, and formulations to various conversational contexts. According to sociolinguists~\citep{conrad2009register}, human linguistic choices are not arbitrary but are closely tailored by speakers to convey not just the informational payload, but also a host of subtle but important social cues~\cite{kilgarriff2005language, kamberelis1995genre}. This concept is called \textit{register}, and it has emerged as one of the most important predictors of linguistic variation in human-human communication~\cite{biber2012register}. Although the relevance of register in human-human communication has been well-established in the sociolinguistic community, the potential applicability of these insights has not yet been explored in the context of chatbot language design.

\citet{argamon2019register} suggests that the sociolinguistic concept of register could be formalized to provide a theoretical basis for machine language generation. To achieve that, chatbots would need to be enriched with computational models that can evaluate the conversational situation and adapt the chatbot's linguistic choices to conform with the expected register, which is similar to the subconscious humans' language production process. The first step toward this goal is to understand whether register theory applies to chatbot interactions and how it might inform chatbots' designs. The cornerstone for this effort is an analysis to expose the perceived effect of chatbot language choices, while reducing the effect of variables other than language (e.g., variations in conversation context and content). This requires parallel conversations that are similar in content, but each represent a different linguistic register.

In our previous work~\cite{chaves2019chatting, chaves2019identifying}, we explored language variation in the context of tourism-related interactions, aiming to expose the relationships between register and the interactional situations within this domain. Our results conformed with established sociolinguistic theories: core linguistic features vary as the situational parameters vary, resulting in different language patterns. In this paper, we move to the next step, analyzing the extent to which the register differences we previously identified play a role in shaping the user's perception of the human-chatbot interaction. The question guiding our investigation is: \textit{Does the use of register-specific language influence the users' perceptions of tourist assistant chatbots?} 

To answer this question, we analyzed the register of two corpora of conversations ($FLG$ and $DailyDialog$) to characterize their register differences and used the outcomes to produce a parallel corpus ($FLG_mod$) in the tourism domain, which have similar informational content as $FLG$, but vary in language use patterns. Then, we performed a user study where we asked participants to compare answers from the two corpora and express their preferences in terms of appropriateness, credibility, and user experience. Finally, we conducted a statistical analysis of our user study data using a Generalized Linear Model~\citep{friedman2010regularization} (binary response) to identify associations between the frequency of core linguistic features and user preferences with respect to language use.

Our results showed that there is an association between linguistic features and user's perceptions of appropriateness, credibility, and user experience, and that register is a stronger predictor of this association than other variables of individual biases (participants, their social orientation, and answers' authors). These outcomes have important implications for the design of chatbots, e.g., the need to design chatbots to generate register-specific language for hard-coded and dynamically generated utterances to improve acceptance and users' perceptions of chatbot interactions.

\section{Background}
\label{sec:related-work}

In the following, we summarize the literature on chatbot language design and the rationale for applying sociolinguistic register analysis as the theoretical foundation for this research. We also discuss our choice of the tourism domain, specifically information search interactions in tourism contexts, as a testbed for our research. 

\subsection{Chatbot Language Design}

Chatbots are typically designed to mimic the social roles usually associated with a human conversational partner, for example, a buddy~\cite{thies2017how}, a tutor~\cite{tegos2016investigation, dyke2013towards}, healthcare provider~\cite{montenegro2019survey, fitzpatrick2017delivering}, a salesperson~\cite{gnewuch2017towards, zhu2018lingke}, a hotel concierge~\cite{lasek2013chatbots}, or, as in this research, a tourist assistant~\cite{chaves2018single, chaves2019chatting}. Research on mind perception theory~\cite{lee2019virtual, heyselaar2019using, keijsers2018mindless} suggests that although artificial agents are presumed to have sub-standard intelligence, people still apply certain social stereotypes to them. It is reasonable, then, to assume that ``machines may be treated differently when attributed with higher-order minds''~\cite{lee2019virtual}. As chatbots enrich their communication and social skills, the user expectations will likely grow as the conversational competence and perceived social role of chatbots approach the human profiles they aim to represent. A variety of factors influence how people perceive chatbot communication skills~\cite{chaves2020how, feine2019taxonomy, tariverdiyeva2019chatbots} and, as user expectations of proficiency increase, one important way to enhance chatbot interactions is by carefully planning their use of language~\cite{kirakowski2009establishing, go2019humanizing}.

Most linguists agree that the language choices made by humans are systematic~\cite{kilgarriff2005language}, and previous research has provided ample evidence that variation within a language can often be accounted for by factors such as individual author/speaker style (e.g.,~\cite{argamon1998routing, leech2007style}), dialect (e.g.~\cite{labov2005atlas, szmrecsanyi2011corpus}), genre (e.g.,~\cite{kamberelis1995genre, paltridge1994genre}), and register (e.g.,~\cite{biber1988variation, conrad2009register}). Among these factors, style has particularly captured the attention of researchers on conversational agents~\cite{feine2019taxonomy, thomas2018style, niederhoffer2002linguistic, jakic2017impact, lin2017stylistic}, with explorations ranging from consistently mimicking the style of a particular character \cite{lin2017stylistic, syed2020adapting} to dynamically matching the style to the conversational partner \cite{niederhoffer2002linguistic, hoegen2019end}.

A number of studies have sought to empirically evaluate the influence of conversational style on user experiences with chatbots~\cite{elsholz2019exploring, araujo2018living}. For example, \citet{elsholz2019exploring} compared interactions with chatbots that use modern English to those that use a Shakespearean language style. Users perceived the chatbot that used the modern English style as easy to use, while the chatbot that used Shakespearean English was seen as more fun to use. \citet{araujo2018living} evaluated the influence of anthropomorphic design cues on users' perceptions of companies represented by a chatbot, where perceptions include attitudes, satisfaction, and emotional connection with the company; one cue, for instance, was the use of an informal language style. Results showed that anthropomorphic cues resulted in significantly higher scores for adjectives like likeable, sociable, friendly, and personal in user evaluations of the interactions (though the relative impact of individual anthropomorphic cues on the outcomes was not evaluated). Similarly, based on an exploratory analysis, \citet{tariverdiyeva2019chatbots} concluded that ``appropriate degrees of formality'' (renamed ``appropriate language style'' in a subsequent work~\cite{balaji2019assessing}) directly correlates with user satisfaction. We note that these studies define ``appropriate language'' as the ``ability of the chatbot to use appropriate language style \textit{for the context}.'' This linkage of perceived appropriateness of language to context is important and reflects clear evidence that appropriateness of language is not absolute, but rather influenced by the user's specific expectations concerning the chatbot's communicative behavior and the stereotypes of the social category~\cite{jakic2017impact, krauss1998language}. For example, when assessing the effects of language style on brand trust in online interactions with customers, \citet{jakic2017impact} concluded that the perceived language fit between the brand and the product/service category increases the quality of interaction. Proficiency in human-like language style may also influence the users' perceptions of chatbot credibility. \citet{jenkins2007analysis} observed that chatbots are deemed sub-standard when users see them ``\textit{acting as a machine}''; similarly, in analyzing the naturalness of chatbots, \citet{morrissey2013realness} found that correct language usage was a determinant in perceived chatbot quality. The failure to convey linguistic expertise compromises credibility~\cite{zumstein2017chatbots}, i.e., the chatbot's ability to convey believability and competence~\cite{sweeney2008effects, mack2008believe}.

Although some scholars define style as ``the meaningful deployment of language variation in a message''~\cite{feine2019taxonomy}, sociolinguistics define style as a set of linguistic variants that reflect aesthetic preferences, usually associated with particular speakers or historical periods \cite{conrad2009register} (e.g., Shakespearean vs. modern English). Sociolinguistic studies also emphasize that the \textit{``core linguistic features like pronouns and verbs are functional''} rather than aesthetic~\cite{conrad2009register}, which points to register. Register theory states that for each interactional situation there is a subset of norms and expectations for using language to accomplish communicative functions~\cite{conrad2009register}. In a conversation, every utterance is influenced by the social atmosphere~\cite{bakhtin2010speech, jabri2008reconsidering}, which is represented in the form of situational parameters, such as the relationship between participants, the purpose of the interaction, and the topic of the conversation~\cite{kamberelis1995genre, conrad2009register}. This results in the emergence of situationally-defined language varieties, which ultimately determine the interlocutor's linguistic choices~\cite{biber2012register, conrad2009register}.

Although the relevance of register in human-human communication has been extensively underlined~\cite{biber2012register}, the extent to which this theory applies to human-chatbot interactions has yet to be widely investigated. There is some evidence suggesting that chatbots should use language appropriate to the service category that the chatbot represents~\cite{balaji2019assessing}. Still, there has been no systematic analysis of how users' perceptions might be influenced by expectations regarding chatbot language, or exploration of specific core linguistic features that determine the appropriateness of language fit. In the next section, we focus on how the register theory applies to human-human communication and why it should be considered in chatbot interactions.

\subsection{Register and Linguistic Variation}

Register theory states that for each interactional situation there is a subset of norms and expectations for using language to accomplish communicative functions~\cite{conrad2009register}; every utterance is influenced by the social atmosphere~\cite{bakhtin2010speech, jabri2008reconsidering}, which is represented in the form of situational parameters, such as the relationship between participants, the purpose of the interaction, and the topic of the conversation~\cite{kamberelis1995genre, conrad2009register}. The influence of these parameters results in the emergence of situationally-defined language varieties~\cite{biber2012register, conrad2009register}. Hence, the register can be interpreted as the distribution of the \textit{linguistic features} in a conversation, given the \textit{context}; the linguistic features consist of the set of words or grammatical characteristics that occur in the conversation, and the context consists of a set of situational parameters that characterize the situation in which the conversation occurs, e.g., the participants, the channel, the production circumstances, and so on.

Several recent studies have shown that register is crucial for linguistic research on language variation and use: most linguistic features are adopted in different ways and to varying extents across different registers. For example, some studies have focused on describing register variation in the use of a narrower set of features, such as grammatical complexity features~\cite{biber2011should, biber2010challenging}, lexical bundles~\cite{biber2004if, hyland2012bundles}, and evaluation~\cite{hunston2010corpus}. The \textit{Longman Grammar of Spoken and Written English}~\cite{biber1999longman} documents the systematic patterns of register variation for most grammatical features in English, exposing the power of register as a significant predictor of linguistic variation. Indeed, we draw on this grammar to support our discussion surrounding the use of particular linguistic features in the context of tourist assistant discourse (see Section~\ref{sec:discussion}).

The value of register for understanding conversational structure is further emphasized by studies showing that \textit{failing} to account for register in linguistic analyses and computational language models can, and often does, result in incorrect conclusions about language use. ~\citet{biber2012register}, for instance, offers many examples of how failing to account for conversational register in a linguistic analysis can result in faulty conclusions. 

Given the crucial role of register in shaping human-human communication, we suggest that register must be accounted for in the design of chatbots language; users' perception of chatbots as competent and trustworthy conversational partners depends on the chatbot's correct use of register. This paper works to provide a practical cornerstone for this broad endeavor by exposing associations between the frequency of core linguistic features (which comprise the conversational register) and user evaluations of the quality of the chatbot interactions. We focus our study on the domain of ``tourist information search'' to (a) analyze the linguistic features relevant to characterizing conversational register in this domain, and (b) show how adherence or failure to adhere to that register impacts user perceptions of conversational quality. To identify the typical patterns of language present in the interactions, we applied register analysis, as developed by~\citet{conrad2009register}, consisting of two main steps. First, \textit{situational analysis} aims to characterize the interactions in the target corpus using a conversational taxonomy based around seven situational parameters: participants, relationship, channel, production, setting, purpose, and topic. Second, \textit{register characterization} analyzes and aggregates the results of the situational analysis to yield a statistical characterization of the linguistic features typically used in domain interactions (in this case, within the tourist assistant domain); the result is a register model for the domain, i.e., a concrete representation of the appropriate register for the given domain.  

\subsection{Chatbots in the tourism domain}

Tourism is one of the fastest-growing economic sectors in the world and is a major category of international trade in services~\cite{unwto2017tourism}. As the sector grows, the demand for timely and accurate information on destinations also increases; a recent survey~\cite{loo2017future} revealed that the Internet is the top source for travel planning. As the penetration of smartphone devices with data access has increased, travelers increasingly search for information and make decisions en-route~\cite{buhalis2011tourism, google2016how, wang2016smartphone}. However, conducting en-route travel information searches using small-screen mobile devices can be an overwhelming experience~\cite{lang2000effect, pawlowska2016tourists}, due to the information overload compounded by a lack of reliable mechanisms for finding accurate, trustworthy, and relevant information~\cite{lang2000effect}. 

\citet{radlinski2017theoretical} suggest that complex information searches could benefit from conversational interfaces and, indeed, several conversational agents with a wide range of characteristics have been developed to improve tourism information search and travel planning~\cite{alexis2017r, ivanov2017adoption, lee2010receptionist, linden1997interactive}. \citet{loo2017future} claims that over one in three travelers across countries are interested in using digital assistants to research or book travel and that travel-related searches for ``tonight'' and ``today'' have grown over 150\% on mobile devices in just two years. In particular, a report on the chatbot market~\cite{grandviewresearch2017chatbot} places the Travel \& Tourism sector as one of the top five markets with the best revenue prospects by 2025. In response to this trend, the number of chatbots within the online tourism sector has increased. The BotList website, for example, lists nearly a hundred available chatbots under the travel category; some examples include the Expedia\footnote{https://botlist.co/bots/expedia} and Marina Alterra\footnote{https://botlist.co/bots/marina-alterra} virtual assistants. 

Travel planning is also a domain in which perceived competence and trustworthiness is central to user experience; advice that is not appropriately presented is unlikely to be trusted and utilized. From a more pragmatic perspective, the popularity of tourist assistants (both human and chatbot) means that a growing corpus of conversations in this domain exists and can serve as the seed for our analysis.

In sum, we selected the tourism advising domain for developing a practical framework for including register in the design of chatbot conversational engines because proper use of conversational register is likely to be particularly critical for user experience in this domain, there is a real-world demand for chatbots for travel advice, and several corpora of human and chatbot interactions in this domain are available.

\section{Research Method}
\label{sec:method}

As aforementioned, this research aimed to explore the extent to which user experience (in terms of perceived appropriateness, credibility, and overall user experience) is related to the conversational register used by a chatbot. For this purpose, we compared conversations expressed in different registers, presenting them to users for evaluation. To isolate the effect of register on perceived user experience, we compared conversations that are equivalent in content but vary in language patterns.

Finding such parallel data–-natural language texts that have the same semantic content, but are expressed in different forms~\cite{nevill1992compression}–-is difficult. Previous studies requiring such parallel data have typically used written texts with multiple versions, e.g., versions of the bible or Shakespearean texts in the original and modern language forms (see~\cite{tikhonov2018wrong}). Although perhaps useful for analysis in an abstract context of NLP research, these corpora portray archaic language centered around topics not likely to be relevant to most modern chatbot users, much less in the design of tourist assistants.

Our approach, therefore, was based on the production of a parallel corpus. This corpus was based on actual conversations, which were carefully manipulated based on register theory to produce conversations of equivalent content, but in differing registers. Unlike previous studies that focus on style~\cite{elsholz2019exploring, hoegen2019end, tariverdiyeva2019chatbots} (i.e., preferences associated with authors or historical period), we relied on the register theory to identify and reproduce language variations that would be plausible for a tourist assistant chatbot to use. Moreover, developing a concrete basis for explicitly manipulating conversational register in the design of chatbot language requires an explicit characterization of the register. Therefore, we identified a set of linguistic features that together characterize the register and show how varying these features affects user perceptions of conversational quality.

\begin{figure}[ht]
  \centering
  \includegraphics[width=11.5cm]{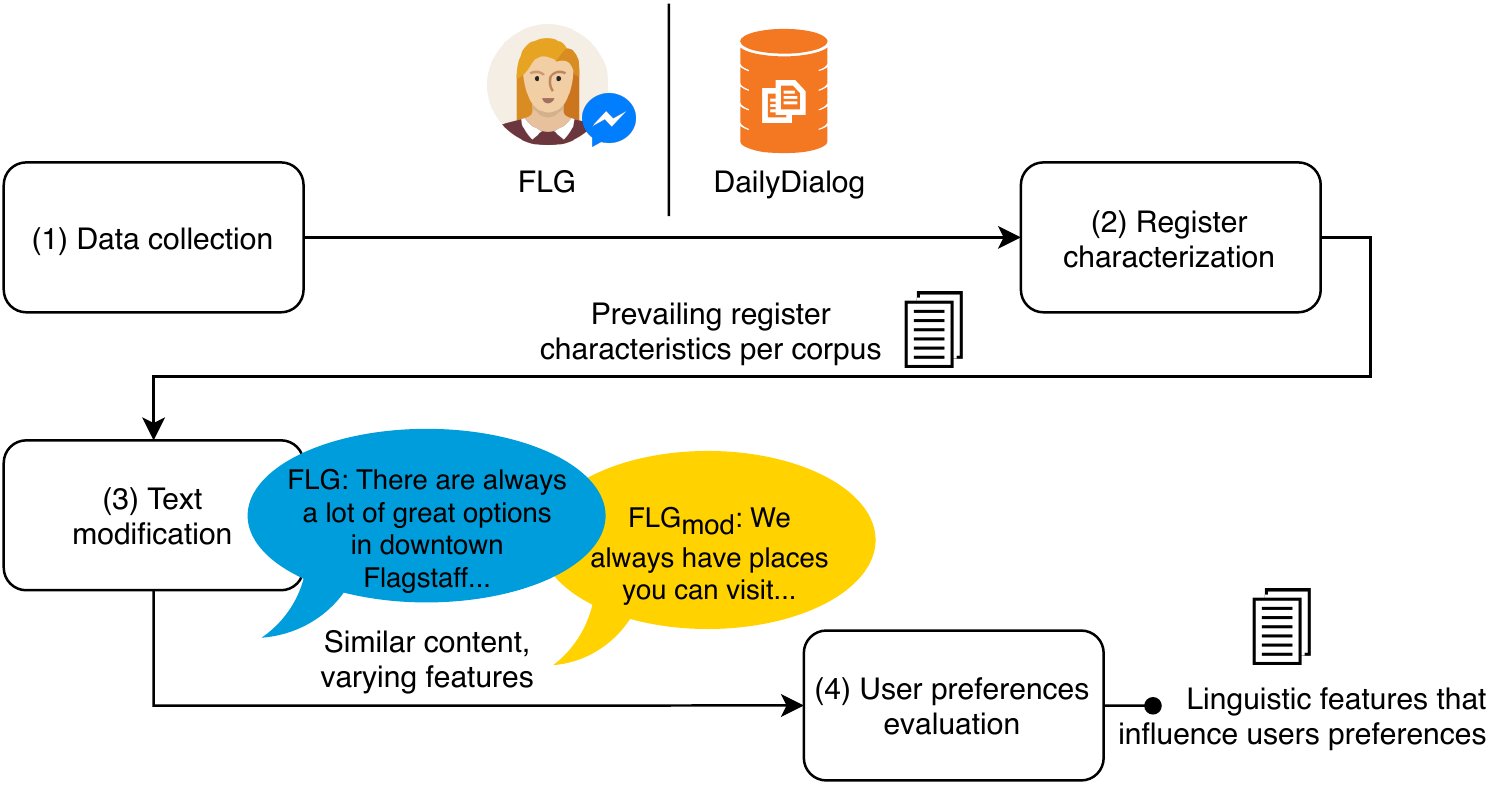}
  \caption{Overview of the research method. The method consists of four main steps, and the outcomes of one step is seeded into the next step.}
  \Description{The figure shows the method steps lined up, in the following order: (1) data collection: the output of this steps is the FLG and DailyDialog corpora; (2) register characterization: the input is the two corpora selected before and the output is the list of prevailing register characteristics per corpus; (3) text modification: the input is the prevailing register characteristics per corpus, and the output is the FLG(mod) corpora, which is a corpus that portrays similar content as FLG, but the patterns of language of DailyDialog. The figure shows the following example: FLG: There are always a lot of great options in downtown Flagstaff...) vs. FLG mod: We always have places you can visit...; (4) user preferences evaluation: the input is the FLG and FLG mod, and the output is the set of linguistic features that influence users preferences.}
  \label{fig:method}
\end{figure}

Our approach comprised four steps, as illustrated in Figure~\ref{fig:method}. We collected two corpora of conversations in the tourism domain and used them to develop parallel corpora that, while equivalent in content, differ across varying dimensions of conversational register. These conversations were then presented to users to generate a multi-faceted evaluation of subjective conversational quality. Finally, the user perceptions were analyzed with respect to the variations in register to expose a relationship between language patterns and user perceptions. These steps are further described below.

\begin{enumerate}
    \item \textbf{Data collection:} to provide a foundation for our analysis, we collected conversations of human domain experts (tourist assistants) interacting with tourists in a text-based tourist information search scenario; we refer this as the $FLG$ corpus. Because conversational register is characterized by comparing linguistic expression in varying interactional situations, we also selected another corpus of conversations in the tourism domain that is available online and is commonly used in natural language research, namely $DailyDialog$~\cite{li2017dailydialog}. Conversations in this corpus span a random variety of daily-life topics from ordinary life, to politics, health, tourism, and other topics. Details about the corpora collection are provided in Section~\ref{sec:data-collection}.
    
    \item \textbf{Register characterization:} our next aim was to characterize the conversational registers present in our two corpora. Based on a broad set of linguistic features that have been previously identified as relevant for characterizing conversational register~\cite{biber1988variation}, we performed register analysis of the $FLG$ and $DailyDialog$ corpora individually, and then statistically compared the patterns of language between them, similar to the analysis performed by ~\citet{chaves2019identifying}. The register characterization step is detailed in Section~\ref{sec:register-tourism}.
    
    \item \textbf{Text modification:} having identified discrete register variations present in the two corpora, our focus shifted to using these register characterizations to produce a parallel corpus in which conversations had equivalent information content, but used a different linguistic format. Specifically, for every answer provided by a tourist assistant in the $FLG$ corpus, we performed linguistic modifications to produce a new corresponding answer that portrays a language pattern that mimics the register characteristics from the $DailyDialog$ corpus; we call this produced parallel corpus $FLG_{mod}$. To assess whether the modified answers in $FLG_{mod}$ preserved the informational content of the original, we performed a study to validate the text modification. We invited participants to compare the parallel answers in terms of naturalness and content preservation. Section~\ref{sec:modification-validation} details the text modification and validation. After performing these foundational steps, we ended up with two parallel corpora ($FLG$ and $FLG_{mod}$).
    
    \item \textbf{Users preferences evaluation:} after developing the parallel corpora that differ solely in the portrayed register, we performed a study to reveal whether users perceived register variations and, if so, what linguistic variations within a register characterization appear to have the greatest impact on aspects of user experience. Overall, we expected to find a preference for the original answers from the $FLG$ corpus, since these are register-specific language produced by humans. To perform the analysis, we selected a subset of tourist questions and their corresponding answers from both $FLG$ and $FLG_{mod}$. Participants were presented with these individual question-answer exchanges and, for each, were asked to choose which answer they preferred based on three distinct measures of quality: appropriateness, credibility, and overall experience. Then, we fitted a statistical learning model to identify the linguistic features that best predict the users' choices. This study is detailed in Section~\ref{sec:user-study}.
\end{enumerate}

Having established an overview of our study method, the following sections detail each of the steps outlined above.

\section{Data collection}
\label{sec:data-collection}

The data collection comprises two sub-steps: (i) to collect a baseline corpus of human-human conversations between tourist assistants and tourists; and (ii) to select a corpus of conversations in the tourism domain. In the following, we introduce the collected corpora.

\subsubsection{FLG} To collect the $FLG$ corpus, we hired three experienced professionals from the Flagstaff Visitor Center in Flagstaff, Arizona, USA, to answer tourist questions about the city and nearby tourist destinations during summer 2018. The official government website reports that Flagstaff receives over 5 million visitors per year~\cite{flagstaff2019city}, including in-state, out-of-state, and international visitors. According to the 2017-2018 Flagstaff Visitor Survey~\cite{combrink2018flagstaff}, Flagstaff is the central hub for visiting tourist destinations such as Grand Canyon National Park, Arizona Snow Bowl, the Navajo and Hopi reservations, and many other local attractions. Regional tourism is significant as well, with a large number of visitors seeking to escape the heat and crowding of the Phoenix metropolitan area. 

The three tourist assistants were native English speakers, female, had some post-secondary education, and had four or more years of experience as tourist assistants. Two of them were 25-34 years old; the other was in the 35-44 age range. Although they had more than four years of experience in providing tourist information in in-person conversations at the Visitor Center, they had never professionally provided information through an online platform.

To recruit tourists to interact with the tourist assistants, we advertised the free tourist assistant online service in the city of Flagstaff through flyers and intercepted tourists at the Flagstaff Visitor Center in Historic Downtown, directing them to a booth to use the service. About 30 tourists participated in the interactions. We also collected tourism-related questions about Flagstaff from websites such as Quora, Google Maps, and TripAdvisor, and a researcher posted these questions to the tourist assistants. The tourist assistants were unaware of the origin of the questions and thought they were always interacting with real tourists. 

The tourist advising conversations were performed through a Facebook Messenger account~\cite{chaves2018tech} over the summer of 2018. The human tourist assistants participated in the study from our lab. Before the first interaction, the tourist assistants participated in a training session, in which we presented the environment and the tools. During the study, a researcher observed the interactions and took notes on comments made by the tourist assistants. Because we wanted to understand the natural linguistic variation in tourism-related interactions, both tourists and tourist assistants were free to interact according to their needs, interests, and knowledge. No tasks were proposed to the tourists, nor were any scripts provided to the tourist assistants. The textual exchanges were exported from Facebook Messenger and archived to create the FLG corpus; the corpus comprises 144 interactions with about 540 question-answer pairs. To analyze the register of the conversation, we only used the answers from the tourist assistants.

\subsubsection{DailyDialog} The second corpus we selected is $DailyDialog$~\cite{li2017dailydialog}, which is a corpus available online and used as a reference for research on natural language generation in the tourism domain. DailyDialog consists of conversations about daily life crawled from websites for English language learning, with topics ranging from ordinary life to politics, health, and tourism. Similar to someone who would use this corpus to train a chatbot, we filtered the original corpus to select only conversations that were originally labeled as ``tourism'' and show customer-service provider interactions, e.g., hotel guest-concierge, business person-receptionist, tourist-tour guide, etc. We chose $DailyDialog$ because it contains a large set of conversations in the tourism domain and it is likely be used as a baseline model for chatbot conversations~\cite{galitsky2019chatbot}.

We downloaded the $DailyDialog$ corpus from its website\footnote{http://yanran.li/dailydialog}. After filtering to focus on tourism-related interactions, the subset of $DailyDialog$ used in this research comprises 999 interactions. Because we are only interested in the utterances produced by the service providers ($DailyDialog$) and tourist assistants ($FLG$), we edited the conversations to remove the tourists' utterances.

\subsubsection{Corpora characteristics} We followed the situational analytical framework proposed by~\citet{conrad2009register} to identify the situational parameters in which the conversations took place. The main outcome of the situational analysis is presented in~\citet{chaves2019chatting} and summarized in Table~\ref{tab:situational-parameters}.

\begin{table}[ht]
    \footnotesize
    \caption{Situational analysis. Situational parameters are extracted from the situational analytical framework~\cite{conrad2009register}}
    \label{tab:situational-parameters}
    \begin{tabular}{lll}
    \toprule
     \textbf{Situational parameter} & \multicolumn{1}{c}{\textbf{DailyDialog}} & \multicolumn{1}{c}{\textbf{FLG}}\\\midrule
     Participants & Customer and service providers & Tourists and tour guides \\
     Relationship & Role, power, and knowledge relations vary & Tourist-tour guide, the latter owns the knowledge \\ 
     Channel & Human-written, representing face-to-face & Written, instant messaging tool \\
     Production & Planned & Quasi-real-time \\
     Setting & Private, shared time, and mostly physically shared place & Private, shared time, virtually shared place \\
     Purpose & Provide a service or information & Information search \\
     Topic & Varies within the context of tourism & Local information (e.g., activities, attractions) \\
    \bottomrule
\end{tabular}
\end{table}

According to register theory~\cite{conrad2009register}, differences in situational parameters result in varying register; people use different patterns of language depending on the context. $DailyDialog$ presents larger variability in terms of situational parameters (e.g., participants, purpose, and topic) than $FLG$. Given differences in the situational parameters, we expect that the language characteristics in $FLG$ differ from the language characteristics in $DailyDialog$. We investigate this claim in the next section, where we discuss our register characterization analysis, which identifies the linguistic features that determine the register of each corpus and how the typical language varies among different situations in the tourism domain. 

\section{Variations of conversational register within tourism-related interactions}
\label{sec:register-tourism}

To characterize the varying conversational registers used in our two corpora conversations, we performed a register analysis~\cite{biber1988variation}. Register analysis consists of identifying the linguistic features typically used in a corpus, which is based on tagging and counting the linguistic features present in the utterances and interpreting them according to their function in the sentence~\cite{conrad2009register, biber1988variation}. We performed register analysis for both $FLG$ and $DailyDialog$ corpora and then compared the outcomes to identify the variations in language use across corpora; the following subsections present this analysis and its outcomes in detail. 

\subsection{Register Analysis}

\subsubsection{Procedures}

Our register analysis relied on information from the Biber grammatical tagger~\cite{biber2017multidimensional} to identify the linguistic variation present in each corpus. Given a set of texts, this tool tags and counts the linguistic features present in the text, and return the counts normalized per 10,000 words. The tagger also calculates \textit{dimension scores} for each text, which are based on aggregations of subsets of features derived using a multidimensional analysis algorithm. The dimension scores reveal the prevailing characteristics of the register (i.e., the levels of personal involvement, narrative flow, contextual references, persuasion, and formality present in the texts)~\cite{biber1988variation}. Details about the tagger can be found elsewhere~\cite{biber2017multidimensional, biber1988variation}.

We first analyzed the dimension scores to understand the linguistic characteristics and varieties of the discourse in each corpus. Following~\citet{biber1988variation}, we applied a one-way multivariate analysis method (MANOVA) to generate a statistical comparison of the dimension scores across corpora, where the dependent variables are the values of the five dimension scores, and the independent variables are the $DailyDialog$ (control group) and the three tourist assistants from the $FLG$ corpus are $TA1$, $TA2$, and $TA3$ (experimental groups). Each \textit{text} corresponds to one observation in our model, where a text is a set of one or more contiguous sentences produced by an interlocutor (i.e., one answer). Given the significant overall MANOVA test, we also performed a one-way univariate analysis ($df=3, 1139$) for each of the five dimensions to identify the individual dimensions that influence the prevailing register characteristics.

Finally, for each of the 49 individual features used to calculate the dimension scores, we performed an ANOVA statistical test ($df=3, 1139$) where the dependent variables are the frequency of occurrences of a feature normalized per 10,000 words, whereas the independent variables are the two corpora: the control group is $DailyDialog$, and the experimental groups are each of the three tourist assistants from the $FLG$ corpus. All the reported statistics use a 5\% significance level ($\alpha=0.05$).

\subsubsection{Results}

The MANOVA revealed that our three tourist assistants' dimension scores are significantly different from the average $DailyDialog$ discourse ($Wilks=0.92, F=6.23, p<0.0001$). Table~\ref{tab:register-stats} summarizes the univariate analysis per dimension.

\begin{table}[ht]
    \caption{Univariate analysis of dimension scores ($df=3,1139$). For each dimension, the table shows the estimated dimension score $\pm$, the standard error per group ($DailyDialog$, $TA1$, $TA2$, $TA3$), and the corresponding F- and p-values.}
    \label{tab:register-stats}
    \begin{tabular}{lrrrrrr}
        \toprule
        & \textbf{DailyDialog}      & \textbf{TA1}              & \textbf{TA2}                & \textbf{TA3}               & \textbf{F-value} &  \textbf{p-value}                   \\
        \midrule
        Dim. 1: Involvement           & 30.50 $\pm$ 1.10  &  14.73 $\pm$ 5.06 &   5.69 $\pm$ 5.12  &  -5.02 $\pm$ 4.86  &  25.58  &  \textless0.0001 \\
        Dim. 2: Narrative flow        & -4.10 $\pm$ 0.09  &  -4.45 $\pm$ 0.41 &  -4.31 $\pm$ 0.41  &  -4.79 $\pm$ 0.39  &   1.30  &  0.2978 \\
        Dim. 3: Contextual references & -6.28 $\pm$ 0.33  &  -3.33 $\pm$ 1.52 &  -1.75 $\pm$ 1.54  &  -2.65 $\pm$ 1.46  &   5.44  &  0.0010 \\
        Dim. 4: Persuasion            &  2.43 $\pm$ 0.30  &   1.98 $\pm$ 1.40 &  -0.02 $\pm$ 1.41  &   1.93 $\pm$ 1.34  &   1.01  &  0.3877 \\
        Dim. 5: Formality             &  0.36 $\pm$ 0.26  &  -0.81 $\pm$ 1.20 &  -1.70 $\pm$ 1.21  &  -2.10 $\pm$ 1.15  &   2.41  &  0.0658 \\
        \bottomrule
    \end{tabular}
\end{table}

The dimensional analysis revealed that $DailyDialog$ portrays an oral discourse while tourist assistants in $FLG$ are more literal and informational than involved (Dimension 1), which can be explained by both the face-to-face nature of the conversations in $DailyDialog$ and the variation in the participants' role and power. $DailyDialog$ also has a more extreme negative score for contextual references (Dimension 3), which might be explained by the shared space and common ground provided by face-to-face interactions. $DailyDialog$ also has a slightly more formal discourse and elaborated language, although this difference is not significant. Both corpora show a descriptive rather than narrative language (negative estimates for Dimension 2) and slightly persuasive language (positive estimates for Dimension 4).

Since our ultimate goal is to reproduce the patterns of language (i.e. register) of $DailyDialog$ within the conversations of the $FLG$ corpus to produce our parallel corpora, we need to identify not only the main linguistic characteristics present in the discourse (e.g., how involved or persuasive is the discourse), but also how these characteristics emerge in each corpus. Although the dimensional analysis reveals the overall register characterization, it is not sensitive enough to identify the prevailing linguistic features that influence the overall discourse. For example, although Dimension 4 is not significantly different, the prevailing linguistic features that contribute to the dimension score varied across corpora. Thus, we statistically compared the occurrences of every linguistic feature per dimension. The left side of Table~\ref{tab:features_modified} lists the linguistic features that vary significantly between the two original corpora ($FLG$ and $DailyDialog$), as revealed by the ANOVA analysis per feature. The table includes the estimates for $DailyDialog$ (control group) and each tourist assistant in $FLG$ (TA1, TA2, TA3) as well as the F-values. A more complete table that includes the non-significant linguistic features and the p-values per feature is presented in the supplementary materials, which also include a glossary with examples of the features.

\begin{table}[ht]
    \footnotesize
    \caption{ANOVA results for individual features comparison between $DailyDialog$ and both original and modified corpora. The left side of the table presents the estimates and standard error for each independent variable ($DailyDialog$, $TA1$, $TA2$, $TA3$) and the F-value. The right side of the table shows the estimates, standard error, and F-values for the three experimental groups ($TA1_{mod}$, $TA2_{mod}$, $TA3_{mod}$) after modifications being performed ($DailyDialog$ column was omitted in the right side to avoid repetition). All the statistics are calculated with $df= 3, 1139$}
    \label{tab:features_modified}
    \begin{tabular}{lrrrrr|rrrr}
        \toprule
         & \multicolumn{5}{c|}{Estimates$\pm$Std.Err. ($FLG$)}  & \multicolumn{4}{c}{Estimates$\pm$Std.Err. ($FLG_{mod}$)}\\ 
        Features	     & $DailyDialog$	    & TA1	            & TA2	            & TA3	            & F     & $TA1_{mod}$	        & $TA2_{mod}$           & $TA3_{mod}$	        & F \\
\midrule \multicolumn{10}{c}{Dimension 1: personal involvement} \\ \midrule
Private verb             &   14.7   $\pm$ 0.8	& 6.7   $\pm$ 3.5	& 5.9	$\pm$ 3.5	& 5.7   $\pm$ 3.3	& 5.6	&    14.5	$\pm$ 3.5	&   16.8	$\pm$ 3.5	&   14.7	$\pm$ 3.2	&  0.1 \\
That-deletion            &   4.8    $\pm$ 0.4	& 0.7	$\pm$ 1.8	& 0.4	$\pm$ 1.8	& 0.6   $\pm$ 1.7	& 5.0	&     4.1	$\pm$ 1.9	&    4.3	$\pm$ 1.9	&   7.3	    $\pm$ 1.8	&  0.8 \\
Contraction              &   26.7	$\pm$ 1.1	& 11.7	$\pm$ 5.1	& 11.9	$\pm$ 5.1	& 1.6   $\pm$ 4.8	& 13.0	&    30.7	$\pm$ 5.1	&   26.5	$\pm$ 5.1	&   22.4	$\pm$ 4.7	&  0.5 \\
Present verb             &   159.2  $\pm$ 1.7	& 125.9	$\pm$ 7.6	& 119.1 $\pm$ 7.7	& 120.8 $\pm$ 7.3	& 21.6	&    157.1	$\pm$ 7.6	&   153.2	$\pm$ 7.6	&   151.4	$\pm$ 7.0	&  0.6 \\
2nd person pronoun       &   81.8	$\pm$ 1.52	& 41.4	$\pm$ 7.0	& 23.4	$\pm$ 7.1	& 44.5  $\pm$ 6.7	& 38.5	&    68.5	$\pm$ 7.1	&   59.4	$\pm$ 7.0	&   66.4	$\pm$ 6.5	&  5.6 \\
1st person pronoun       &   55.1	$\pm$ 1.3	& 18.2  $\pm$ 5.9	& 16.8	$\pm$ 6.0	& 11.0	$\pm$ 5.7	& 40.6	&    37.6	$\pm$ 6.0	&   39.6	$\pm$ 6.0	&   41.9	$\pm$ 5.5	&  6.1 \\
Causative subord.        &   0.20	$\pm$ 0.06	& 1.19	$\pm$ 0.29	& 0.00	$\pm$ 0.29	& 0.33	$\pm$ 0.28	& 3.93	&    0.39	$\pm$ 0.28	&   0.00	$\pm$ 0.28	&   0.32	$\pm$ 0.26	& 0.39 \\
Discourse particle       &   6.3    $\pm$ 0.5	& 3.7	$\pm$ 2.3	& 0.4	$\pm$ 2.3	& 0.0	$\pm$ 2.2	& 4.9	&    4.3	$\pm$ 2.3	&   3.5     $\pm$ 2.3	&   3.5     $\pm$ 2.1	&  1.2 \\
Indefinite pronoun       &   7.57	$\pm$ 0.53	& 4.79	$\pm$ 2.43	& 1.60	$\pm$ 2.46	& 2.23	$\pm$ 2.33	& 3.68	 &   5.08	$\pm$ 2.47	&   3.46	$\pm$ 2.45	&   3.36	$\pm$ 2.27	& 2.13 \\
Coord. conj. (clause)    &   6.6    $\pm$ 0.4	& 19.0  $\pm$ 1.9	& 8.8	$\pm$ 2.0	& 3.1	$\pm$ 1.9	& 15.0	&    6.4	$\pm$ 1.9	&   6.7     $\pm$ 1.9	&   6.2     $\pm$ 1.8	&  0.0 \\
Final preposition        &   1.50	$\pm$ 0.21	& 4.9   $\pm$ 0.98	& 0.3   $\pm$ 1.0	& 2.0   $\pm$ 1.0	& 4.5	&    1.4	$\pm$ 0.9	&   0.3     $\pm$ 0.9	&   0.8 	$\pm$ 0.9	&  0.8 \\
Nouns                    &   225.6	$\pm$ 2.5	& 277.7	$\pm$ 11.6	& 328.0 $\pm$ 11.8	& 302.1	$\pm$ 11.2	& 41.9	&    258.1	$\pm$ 11.6	&   280.5	$\pm$ 11.4	&   275.6   $\pm$ 10.6	&  15.5 \\
Prepositions             &   70.6	$\pm$ 1.4	& 100.1	$\pm$ 6.6	& 112.1 $\pm$ 6.7	& 91.3	$\pm$ 6.4	& 20.2	&    80.0	$\pm$ 6.6	&   77.2	$\pm$ 6.5	&   69.9	$\pm$ 6.0	&  1.0 \\
Attributive adjective    &   19.5	$\pm$ 0.9	& 43.7  $\pm$ 4.2	& 45.1  $\pm$ 4.2	& 55.1	$\pm$ 4.0	& 43.6	&    26.5	$\pm$ 4.0	&   23.0	$\pm$ 4.0	&   26.3	$\pm$ 3.7	&  2.1 \\
\midrule \multicolumn{10}{c}{Dimension 2: narrative flow} \\ \midrule
3rd person pronoun       &   4.0    $\pm$ 0.5	& 12.7  $\pm$ 2.1	& 9.5   $\pm$ 2.1	& 9.8	$\pm$ 2.0	& 9.7	&    5.0	$\pm$ 2.0	&   5.1     $\pm$ 2.0	&   3.8	    $\pm$ 1.8	&  0.2 \\
\midrule \multicolumn{10}{c}{Dimension 3: contextual reference} \\ \midrule
WH-rel. cl. (subject)    &   0.2    $\pm$ 0.1	& 2.2	$\pm$ 0.3	& 0.1   $\pm$ 0.3	& 0.6	$\pm$ 0.3	& 13.2	&    0.8	$\pm$ 0.3	&   0.0     $\pm$ 0.3	&   0.4	    $\pm$ 0.3	&  1.8 \\
Nominalization           &   21.0	$\pm$ 0.9	& 27.4	$\pm$ 4.2	& 35.4  $\pm$ 4.2	& 23.3	$\pm$ 4.0	& 4.4	&    23.0	$\pm$ 4.1	&   26.2	$\pm$ 4.1	&   21.3	$\pm$ 3.8	&  0.6\\
Time adverbial           &   7.5	$\pm$ 0.5	& 3.5	$\pm$ 2.2	& 1.6   $\pm$ 2.3	& 2.0	$\pm$ 2.2	& 4.8	&    4.5	$\pm$ 2.3	&   4.7 	$\pm$ 2.3	&   9.1	    $\pm$ 2.1	&  1.2 \\
Adverb                   &   36.3	$\pm$ 1.2	& 46.8	$\pm$ 5.3	& 35.2  $\pm$ 5.4	& 23.7	$\pm$ 5.1	& 3.4	&    40.8	$\pm$ 5.3	&   40.9	$\pm$ 5.2	&   33.3	$\pm$ 4.9	&  0.6 \\
\midrule \multicolumn{10}{c}{Dimension 4: persuasiveness} \\ \midrule
Prediction modal         &   21.8	$\pm$ 0.8	& 10.5	$\pm$ 3.7	& 8.4   $\pm$ 3.7	& 14.0	$\pm$ 3.6	& 7.9	&    22.9	$\pm$ 3.8	&   18.1	$\pm$ 3.8	&   19.0    $\pm$ 3.5	&  0.6 \\
Suasive verb             &   0.6    $\pm$ 0.1	& 4.5	$\pm$ 0.7	& 1.1   $\pm$ 0.7	& 3.0	$\pm$ 0.6	& 16.2	&    1.1	$\pm$ 0.6	&   0.8 	$\pm$ 0.6	&   0.3     $\pm$ 0.5	&  0.4 \\
Conditional subord.      &   2.5	$\pm$ 0.3	& 4.0	$\pm$ 1.2	& 6.5   $\pm$ 1.3	& 5.9	$\pm$ 1.2	& 6.0	&    1.7	$\pm$ 1.2	&   2.9	    $\pm$ 1.2	&   2.8	    $\pm$ 1.1	&  0.2 \\
Split auxiliary          &   1.4	$\pm$ 0.2	& 5.5	$\pm$ 0.9	& 1.5   $\pm$ 0.9	& 1.4	$\pm$ 0.9	& 6.3	&    1.9	$\pm$ 0.9	&   1.4	    $\pm$ 0.9	&   1.9     $\pm$ 0.8	&  0.2 \\
\midrule \multicolumn{10}{c}{Dimension 5: formality} \\ \midrule
Adverbial--conj.         &   5.2	$\pm$ 0.4	& 2.5	$\pm$ 1.8	& 1.2   $\pm$ 1.9	& 0.7   $\pm$ 1.8	& 4.0	&    3.1	$\pm$ 1.9	&   4.6     $\pm$ 1.8	&   4.4     $\pm$ 1.7	&  0.5 \\
        \bottomrule
    \end{tabular}
\end{table}

As indicated in Table~\ref{tab:features_modified}, the register analysis reveals 22 linguistic features that vary significantly across corpora through all of the five register dimensions. As we anticipated, differences in the situational parameters influenced the patterns of language observed in the corpora, with the typical language presented in $DailyDialog$ varying significantly from that presented in $FLG$ for a core set of linguistic features. 

\section{Text modification}
\label{sec:modification-validation}

Having characterized the differences in register between the $FLG$ and $DailyDialog$ corpora, our next step was to clone the $FLG$ corpus and then use the register characterization to modify its utterances, mimicking the register characteristics observed in the $DailyDialog$ corpus. In psycholinguistics, linguistic modification consists of changing the language of a text while preserving the text's content and integrity, which includes using familiar or frequently used words~\cite{sato2007guide, bosher2008effects, long1993modifications}. We applied this technique to alter the linguistic features in $FLG$ to approximate its language to the patterns presented in $DailyDialog$. The new corpus, which we call $FLG_{mod}$, is paired with the original $FLG$ corpus to form a pair of parallel corpora equivalent in topic, participants, and informational content, but expressed in varying patterns of language.

One important aspect of text modification is that both the informational content and basic linguistic integrity of the text should be preserved~\citep{sato2007guide, bosher2008effects, long1993modifications}, otherwise, the quality of the produced sentences can be compromised. Therefore, we performed a validation study where participants proof-read the original and modified answers and assessed the modifications in terms of content preservation and quality attributes, such as naturalness and meaningfulness.  
\subsection{Modification process}

We manipulated the answers in $FLG$ to approximate the estimate values (presented in Table~\ref{tab:features_modified}) of a particular feature to the corresponding estimate in $DailyDialog$. For example, the estimate for \textit{private verbs} in $DailyDialog$ is $14.70$ occurrences per 10,000 words, whereas in $FLG$ the greatest rates for \textit{private verbs} is $7.29$; therefore, we want to increase the occurrences of \textit{private verbs}, until we reach an estimate that is closer, and not significantly different from $14.70$ for every tourist assistant in $FLG$ corpus.

\subsubsection{Procedures}

Inspired by previous studies on chatbot language use (see, e.g., ~\cite{elsholz2019exploring}), modifications were performed semi-manually, using the AntConc tool~\cite{anthony2005antconc} and a Python script as support tools. The Python script took in a list of paired inputs, where the first parameter is a text present in the original data that needs to be changed (target); and the second parameter is the text that will replace the original (goal). The script then searched for all the occurrences of the first parameter in the $FLG$ corpus and replaced it with the second. This process is repeated until there are no more entries in the list. The AntConc tool is used to help identify the targets to be added to the Python input list. The script then saves the modified interactions and continues to generate the complete $FLG_{mod}$ corpus. The new texts were then submitted to the Biber tagger and subjected to renewed register analysis, as described in Section~\ref{sec:register-tourism}. If the features of interest in $FLG_{mod}$ were still statistically different from the estimates in $DailyDialog$, the process was repeated until the cumulative changes gathered in the modified corpus yielded a register analysis result that was as similar as possible to the register profile of $DailyDialog$. 

\subsubsection{Results}

The right side of Table~\ref{tab:features_modified} shows the comparison between $DailyDialog$ and $FLG_{mod}$. The table shows the estimates for each tourist assistant in $FLG_{mod}$ and the F-value compared to the $DailyDialog$ (control group). For three features, namely \textit{nouns}, \textit{first-person pronouns}, and \textit{second-person pronouns}, linguistic modification did not reach a non-significant difference, although we substantially reduced the F-values. The problem with these features is that differences between $DailyDialog$ and $FLG$ were so extreme that forcing them to non-significant levels in the modified corpus could affect the distribution of the co-occurring features (e.g., increasing verbs associated with pronouns) or produce artificial changes that could harm the content preservation and the naturalness of the resulting answer. For all the other features, the modifications reached non-significant differences. Table~\ref{tab:example-modification} shows an example of a modified answer. Although both the counts for the co-occurring features and the length of the answers (which influences the normalized counts) changed due to the modifications, features that were not statistically significant when compared to $FLG$ were still not significant when compared to $FLG_{mod}$. The statistical results for the non-significant features can be found in the supplementary materials.

\begin{table}[ht]
    \centering
    \small
    \caption{Example of a modified answer. The left side shows the answer provided by a tourist assistant in the original data collection. The right side shows the corresponding answer, modified to portray features that mimics $DailyDialog$ linguistic form. Modified words are highlighted in bold and the tags attributed to the words are between square brackets.}
    \label{tab:example-modification}
    \begin{tabular}{p{7cm}p{7cm}}
    \toprule
        \multicolumn{1}{c}{\textbf{Original answer from $FLG$ corpora}} & \multicolumn{1}{c}{\textbf{Modified answer ($FLG$ content, $DailyDialog$ form)}}\\
    \bottomrule
    \toprule
       Well there are always a lot \textbf{of} \textit{[preposition]} \textbf{great options} \textit{[attributive adjective, noun]} \textbf{in downtown Flagstaff} \textit{[preposition, nouns]} \textbf{for} \textit{[preposition]} live music at bars like The State Bar and the Hotel Monte Vista. You can see what \textbf{activities} \textit{[nominalization]} are going \textbf{on} \textit{[preposition]} at www.flaglive.com \textbf{for music and events} \textit{[preposition, nouns]}.
       & Well \textbf{we always have} \textit{[1st person pronoun, present verb]} places \textbf{you} can \textbf{visit} \textit{[2nd person pronoun, present verb]} that \textbf{have} \textit{[present verb]} live music. \textbf{I'd suggest} \textit{[1st person pronoun, contraction, prediction modal, suasive verb]} bars like The State Bar and the Hotel Monte Vista. You can see what music and events are happening at www.flaglive.com.\\
    \bottomrule
    \end{tabular}
\end{table}

In summary, for each question-answer pair in the original $FLG$, there is a corresponding question-answer in $FLG_{mod}$, where the answer has equivalent informational content but is expressed in a different register. The patterns of language use in $FLG_{mod}$ mimic those in $DailyDialog$, which is, on average, more personally involved and oral, with additional features for persuasion and formality. To evaluate the quality of the modifications, we asked human subjects to compare the content in the original and modified versions of the answers and assess the naturalness of the modified answers, as detailed in the following section.

\subsection{Validation of Modifications}

We performed a validation study to verify whether the modifications preserved the content and the naturalness of the original text. Details of this modification process are presented in the following subsections.

\subsubsection{Procedures}

\begin{figure}[ht]
  \centering
  \includegraphics[width=10cm]{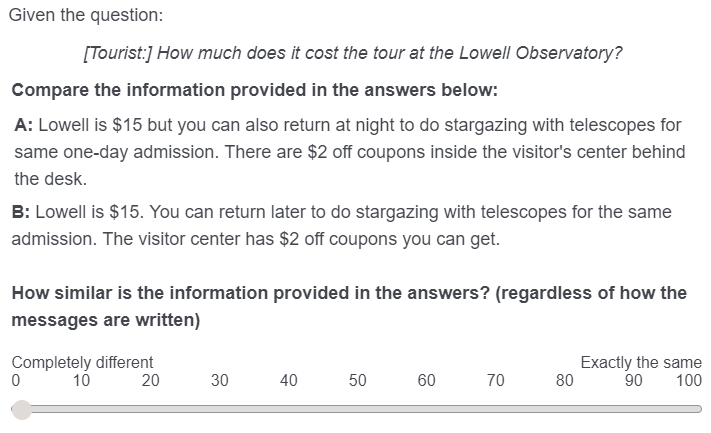}
  \caption{Example of a content preservation question. Participants selected their answer to the question using the slider, where 0 represents that the content in the answers is completely different, while 100 represents that the content in the answers A and B is exactly the same.}
  \label{fig:content-preservation}
  \Description{A figure that shows an example of a content preservation question. The questions say the following: "Given the tourist question: How much does it cost the tour at the Lowell Observatory? Compare the information provided in the answers below:" Option A: "Lowell is 15 US dollars but you can also return at night to do stargazing with telescopes for same one-day admission. There are 2 US dollars coupons inside the visitor's center behind the desk." Option B: "Lowell is 15 US dollars. You can return later to do stargazing with telescopes for the same admission. The visitor center has 2 US dollars coupons you can get." How similar is the information provided in the answers? (regardless of how the messages are written). There is a slider numbered from 0 to 100, where the 0 is labeled as "completely different" and the 100 is labeled as "Exactly the same." Participants would select their answer in the slide.}
\end{figure}

We randomly selected 54 (10\%) question-answer pairs from the parallel corpora and asked participants to judge the content preservation and naturalness of the answers. We collected data via an online questionnaire, where each participant assessed two blocks of questions, presented in random order. In one block, participants were presented with a tourist question and two possible answers (A and B, respectively), which correspond to the original answer from $FLG$ and modified version from $FLG_{mod}$. Participants were unaware of how the answers A and B were produced. For each question-answer pair presented on the screen, participants were invited to rate how similar the information provided in the answers was, regardless of how the messages were written. Participants used a slider to select the similarity level, where the extremities of the sliders were labeled with ``Completely different'' (0) and ``Exactly the same'' (100). One example of a content preservation question is presented in Figure~\ref{fig:content-preservation}. Each participant assessed 10 randomly selected question-answer pairs for content preservation.

In the second block of the assessment, we were interested in the naturalness of the answers. We selected a subset of 27 question-answer pairs from the original set of 54. Participants were presented with a question and one single answer, either from $FLG$ or $FLG_{mod}$, at a time. Considering the question-answer on the screen, participants were invited to use a seven-point Likert scale (1:completely disagree/7:completely agree) to rate the answer on four dimensions: natural, complete, meaningful, and well-written. Each participant rated nine question-answer pairs randomly selected from the 54 possible (27 answers from $FLG$ and 27 answers from $FLG_{mod}$), as well as one attention check.

To evaluate the content preservation, we fitted an intercept-only, mixed effect linear model~\cite{bates2015fitting} with the score for content similarity as the dependent variable. The random effects are the questions and the participants' identification. We considered the content preservation for a modification to be reasonable (i.e., content essentially the same) if the estimate for the intercept (taking into account the standard error) stayed above the upper quartile ($\theta\pm\sigma > 75$).

The naturalness was evaluated using a Cumulative Link Mixed Model (CLMM) for ordinal data~\cite{christensen2019ordinal}. The four rated items were individually assessed, and the ratings per item were combined into negative ratings (1-3), neutral (4), and positive ratings (5-7). We fitted a model with the rates for each item as the dependent variable, the corpora as the independent variable (original vs. modified) and three random effects, namely the participants, the questions, and the items per question.

\subsubsection{Participants}

Participants were recruited through Prolific\footnote{https://www.prolific.co}, in February 2020. Prolific is an online recruitment service explicitly designed for the scientific community to enable large-scale recruitment of willing research participants (see~\cite{palan2018prolific}). We recruited a total of 90 participants, but two were later discarded due to failure to answer the attention check ($N=88$). Most participants were female (56), and the age range was 18-77 ($\mu=33.6$ years-old, $\sigma=11.3$). All the participants claimed English as their first language and were located in USA territory. Each participant received \$5 (US dollars) as a compensation for their participation.

\subsubsection{Results}

Our content preservation dataset contains 880 observations (10 observations per participant). Each question-answer pair was evaluated from 14 to 18 times.

The model shows that the estimated mean for content similarity is $86.78$ (SE=$1.39$, df=$99.5$), and the confidence interval is $(84, 89.5)$, which is reasonably above the upper quartile ($75$). Both random effects are significant, and the effect of participants has the largest variance (see Table~\ref{tab:random-effects}). Hence, we conclude that the linguistic modifications made in producing the $FLG_{mod}$ corpus reasonably preserved the content of answers in $FLG$.

\begin{table}[ht]
    \centering
    \caption{Random effects. The variance explained by the participants is larger than the residual variance, which shows that a significant portion of the variation is influenced by participants' biases.}
    \label{tab:random-effects}
    \begin{tabular}{lrr}
        \toprule
        Groups                  & Variance    & Std.Dev.    \\
        \midrule
        PID (Intercept)         & 143.674     & 11.986      \\
        Question (Intercept)    & 9.112       & 3.019       \\
        Residual                & 102.486     & 10.124      \\
        \bottomrule
    \end{tabular}
\end{table}

Regarding naturalness, the number of positive ratings is consistently higher than negative ratings for both corpora (see Table~\ref{tab:ratings_naturalness}). The CLMM models show that the scores for every item do not significantly vary, as presented in Table~\ref{tab:results_naturalness}. Additionally, the participants' biases account for a lot of variance (see supplementary materials for random intercepts results). Hence, we conclude that the answers in $FLG_{mod}$ are not significantly different in terms of naturalness from the answers in $FLG$.

\begin{table}[ht]
    \centering
    \caption{Number of times the original and modified answers received a negative, neutral, and positive scores. Both original and modified answers consistently received more positive than negative scores.}
    \label{tab:ratings_naturalness}
    \begin{tabular}{lrrrr}
        \toprule
        group                   & Negative (1-3)  & Neutral (4) & Positive (5-7)    \\
        \midrule
        Original ($FLG$)        & 180             & 84          & 1317              \\
        Modified ($FLG_{mod}$)  & 255             & 105         & 1210              \\
    \bottomrule
    \end{tabular}
\end{table}

\begin{table}[ht]
    \centering
    \caption{CLMM results per evaluated item. The table shows the estimate, standard error, Z-values, and p-values for each item.}
    \begin{tabular}{lrrrr}
        \toprule
        Item	        & Estimate  & SE        & z         & Pr (>|z|) \\
        \midrule
        Natural	        & -0.61     & 0.32      & -1.90     & 0.06      \\
        Meaningful	    & 0.07      & 0.051     & 1.42      & 0.16      \\
        Complete        & -0.10     & 0.29      & -0.34     & 0.74      \\
        Well-written    & -0.58     & 0.35      & -1.66     & 0.10      \\
        \bottomrule
    \end{tabular}
    \label{tab:results_naturalness}
\end{table}

In summary, our text modification process produced a parallel corpus, $FLG_{mod}$, with equivalent informational content as the $FLG$ corpus, but with the register characteristics of the $DailtDialog$ conversations. Our validation study indicates that the modifications introduced to generate the $FLG_{mod}$ corpus preserved the content and the naturalness expressed in $FLG$.

\section{Register-specific language for chatbots}
\label{sec:user-study}

In our last research step, we finally address the motivating question driving our effort: investigating whether users are sensitive to changes in conversational register and how such differences in register impact perceived quality of the interaction and overall user experience. To explore this issue, we compared original and modified corpora in a study to identify which answers the participants perceive as more appropriate, credible, and providing the best user experience. Considering that human tourist assistants produced the language in $FLG$, and therefore it is likely to be register-appropriate to the proposed interactional situation, we hypothesized that findings would show a preference for answers from $FLG$; the answers from $FLG_{mod}$ should score lower, as we have artificially modified them to introduce a register that is less likely to be appropriate to the situational parameters in $FLG$. Our analysis additionally included variables to represent the individual interlocutors (both assistants and participants) to compare the strength of these variables when compared to the register variation expressed in the corpora.

\subsection{Measurements}

User experience (UX) refers to the overall experience of a person using a software product, which includes their perceptions and attitudes such as emotions, beliefs, behaviors, and accomplishments~\citep{iso2018ergonomics}. Because user experience is a very broad concept, it is crucial to delimit the scope of the term for this particular study. User experience is often measured in terms of usability metrics, such as effectiveness, efficiency, and satisfaction~\citep{mcnamara2006functionality, radziwill2017evaluating, finstad2010usability}\footnote{Effectiveness, efficiency, and satisfaction are part of the general definition of usability, according to the ISO 9241-11~\citep{iso2018ergonomics}. Usability, in turn, is part of the overall users experience and, in many cases, the two terms are used interchangeably~\citep{tullis2013measuring}. In this research, we differentiate usability from user experience, where usability is the ability to carry the task successfully and user experience focuses on the user perceptions and behavior resulting from the interaction~\citep{tullis2013measuring}.}. In this research, however, we controlled for the chatbot's technology and knowledge as well as the user's tasks, since the main goal was to evaluate the user perceptions regarding the varying patterns of language use. Thus, the usability constructs, such as task success, robustness, and ease of use, was equivalent across treatments.

Thus, for this research, user experience is defined in terms of attitudinal metrics. Specifically, we measured specific attributes that are potentially influenced by the user's expectations and perceptions of the chatbot's language use. Since the conversation's participants and the relationship among them are pointed out in the situation analysis framework~\citep{conrad2009register} as characteristics that influence the register, we evaluated whether the register variation influence how \textit{appropriate} is the language, given the chatbot's social role. Appropriateness speaks to the fit between what the user expects in that conversational context in terms of linguistic form and content and what they actually encounter in the answer. We expected this to be influenced by the tourist's expectations concerning the assistant's communicative behavior and the stereotypes of the social category~\cite{jakic2017impact, krauss1998language}. 

The perceived communicative behavior might also influence how the users perceive the chatbot's \textit{credibility}. Credibility is presented as a rating of confidence in the accuracy of the information contained in the answer; the failure to convey expertise through language compromises credibility~\cite{zumstein2017chatbots} and trustworthiness~\cite{sweeney2008effects, mack2008believe}. According to~\citet{corritore2005measuring}, credibility consists of four factors: honesty, expertise, reputation, and predictability. In this study, we focus on the expertise and honesty factors, which represent a chatbot's perceived competence and believability. Since the participants have no previous experiences with the studied chatbot, reputation and predictability factors did not apply.

Finally, the overall user experience represents the user's general satisfaction with the provided answer and, consequently, the interaction's outcome. Considering that \textit{what} is said is similar between $FLG$ and $FLG_{mod}$, user experience should be mostly influenced by \textit{how} it is said. In the following subsections, we present the method used to evaluate user perceptions of the two parallel corpora and its outcomes.

\subsection{Procedures}

We selected 10\% of the question-answer pairs from the parallel corpora to be evaluated. Considering the semi-manual process, in some cases the answers to a given question were very similar between the two corpora, i.e., one particular answer may not have been modified at all as part of our register-shifting process. To focus our comparison on answers with distinct differences in register, we selected answers that had been substantially modified: we calculated the Levenshtein distance~\cite{doan2012levenshtein, wiki2020algorithm} between the pairs of original and modified answers, and selected the 54 question-answer pairs with the highest values of distance. The Levenshtein distance corresponds to the minimum cost (i.e., the number of insertions, deletions, or substitutions) of transforming one text into the other~\cite{doan2012levenshtein}. See the supplementary materials for the evaluated question-answers pairs and the corresponding Levenshtein distance values.

\begin{figure}[ht]
  \centering
  \includegraphics[width=10cm]{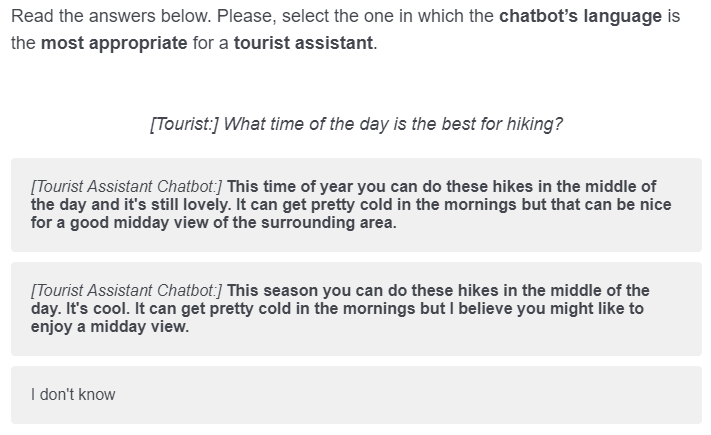}
  \caption{Example of a question from the study. In this example, the participant was invited to select the answer that portrays the most appropriate language. Participants selected their responses by clicking on their preferred answer or on the ``I don't know'' option.}
  \label{fig:example-question}
  \Description{A figure that shows an example of a question from the study. The question says the following: "Read the answers below. Please, select the one in which the chatbot's language is the most appropriate for a tourist assistant." "[Tourist:] What time of the day is the best for hiking?" Following the question presentation, there are three clickable options with the following content: (i) "[Tourist Assistant Chatbot:] This time of year you can do these hikes in the middle of the day, and it's still lovely. It can get pretty cold in the mornings but that can be nice for a midday view of the surrounding area." (ii) "[Tourist Assistant Chatbot:] This season you can do these hikes in the middle of the day. It's cool. It can get pretty cold in the mornings but I believe you might like to enjoy a midday view." (iii) "I don't know. Participants selected their preferred answer by clicking on one of the provided options.}
\end{figure}

We collected participant responses via an online questionnaire. After reading and agreeing with the informed consent, participants were introduced to the task: given a tourist question, identify the answer that best represents a tourist assistant's discourse. For this experiment, participants were told that the answers would be provided by a chatbot. For each tourist's question, presented in the screen one at a time, participants could choose one out of three options: the original answer (from $FLG$), the modified version (from $FLG_{mod}$), and ``I don't know'' (see an example in Figure~\ref{fig:example-question}). Original and modified answers were presented in a randomized order, whereas ``I don't know'' was always the last option on the list. Supplementary materials contain a sample of this instrument for the other constructs as well.

Constructs were also evaluated one at a time. Thus, we first showed a definition of the construct of interest (e.g., appropriateness), and then the ten question-answer pairs to be evaluated for that construct, which were nine question-answer pairs extracted from the corpora and one attention check. In total, each participant evaluated 27 different question-answer pairs (9 per construct, without repetition across constructs) randomly selected from the possible 54. The order of the constructs was also randomized. In the end, participants answered the demographics and social orientation questionnaire. We used the social orientation items proposed by~\citet{liao2016what}; the social orientation toward chatbots determines the participants' preferences regarding human-like social interactions with chatbots~\cite{liao2016what}.

The outcome of the questionnaire consists of the users' preferences regarding which answer (original vs. modified) is the most appropriate and credible for a tourist assistant, as well as the answer that would result in the best user experience.

\subsection{Participants}
\label{sec:participants}

Participants were recruited through Prolific\footnote{https://www.prolific.co}, in March 2020. We received a total of 193 submissions, 15 of which were discarded due to either technical issues in the data collection or failure to answer the attention checks ($N=178$). All the participants claimed English as their first language and were located in the USA. Most participants had either a four-year bachelor's degree (59) or some college, but no degree (49). 21 participants had Master's degree, and the other 21 graduated from high school. Common educational backgrounds were STEM (54), Arts and Humanities (42), and Other (32). Three participants had non-binary gender, 86 declared themselves as female, and 89 as male. The age range is 19-73 ($\mu=33.10$ years-old, $\sigma=11.08$).

168 participants declared that they search for travel information online, but 99 declared that they had never used online assistance when traveling. Only 21 participants had never interacted with chatbots before, and 82 said they had interacted five or more times. The distributions of participants are detailed in the supplementary materials. Participants could provide comments and feedback on the studies through a message resource provided by the platform. Only two participants used that resource to report a problem when loading the study page and to provide positive feedback on the completed study.

In order to augment and clarify our quantitative findings, we invited 12 participants from the Northern Arizona University community (students and staff) to perform the study in the lab and share the reasoning about their choices. At least two researchers sat with participants in the lab during these sessions and independently took notes on participants' comments. After every session, the researchers debriefed about their notes and merged the outcomes. Since the number of lab participants is limited, we did not use the qualitative data to draw conclusions regarding the participant's preferences. Instead, we used these notes only to support the discussion presented in Section~\ref{sec:discussion}, pointing out examples of quotes from participants that align with our quantitative results. Quotes are identified with the notation [LabPn], where ``LabP'' stands for ``Lab Participant'' and ``n'' is a number between 1 and 12 that represents the participant. The 12 answers to the questionnaire were not included in the statistical analysis to avoid data source bias. Each participant (for both Prolific and lab studies) received \$5 (US dollars) as a compensation for their participation.

\subsection{Analysis of the linguistic features}
\label{sec:linguistic-features-analysis}

To model the users' preferences between original and modified versions of $FLG$, we fitted a generalized linear model (GLM) for two-class logistic regression, using the $glmnet$ package in R~\cite{friedman2010regularization}. Because we are interested in the difference between original and modified versions of $FLG$ for each linguistic feature of interest (listed in Table~\ref{tab:features_modified}), we calculated the $\text{original} - \text{modified}$ counts, and input this difference into the model. For example, suppose that one particular answer provided by the tourist assistants in the original $FLG$ corpus has $31.3$ \textit{private verbs} (normalized per 10K words), and that after linguistic modification, the corresponding answer has $62.5$ \textit{private verbs}. Thus, for this particular answer, the value for \textit{private verbs} input into the model is $31.3-62.5=-31.2$. A negative value means that the occurrences of that feature were increased in the modification process for that particular answer. In contrast, a positive value for a feature means that the occurrences of that feature were reduced in the modification process.

\subsubsection{Problem Definition}
\label{sec:problem-definition}


Consider a model with the response variable $Y=\{0, 1\}$ where $0$ represents the original answers ($FLG$) and $1$ represents the modified answers. The L1-regularized logistic regression algorithm models class-conditional probabilities through a linear function of the predictors~\cite{friedman2010regularization}. The prediction function is defined as:

$$f(x)= w^T x + \beta$$

where $x$ is a feature vector of $p$ real numbers representing (i) the difference between original and modified counts per linguistic feature; (ii) variables representing the participant who answered the question ($1$ if the observation was answered by that participant, $0$ otherwise), the participants' self-assessed social orientation ($1$ to $7$), and the author of the answer ($1$ if the answer was authored by that tourist assistant, $0$ otherwise). We want to learn a $p$ vector $w$ of weights and a real scalar intercept $\beta$. The L1-regularization ensures that the learned model has a sparse/interpretable $w$ (some entries will be exactly zero; these entries correspond to features that are not used/important for prediction). The prediction function $f(x)$ gives real-valued predictions for the given feature vector $x$. The logistic link function that finds the predicted probability in $[0,1]$ is

$$p(x) = \frac{1}{1+\exp(-f(x))}$$

The function predicts the negative answer (i.e., 0:original) when $f(x)<0$ and $p(x)<0.5$, while the positive answer (i.e., 1:modification) is predicted when $f(x)>0$ and $p(x)>0.5$. 
For comparison purposes (to determine an upper bound on prediction accuracy), we fitted two non-linear learning models: random forest and gradient boosting. For the random forest, we used the \texttt{party} package in R, which provides an implementation of random forests with conditional inference trees~\cite{hothorn2006survival, strobl2007bias, strobl2008conditional}. For the gradient boosting, we used the \texttt{xgboost} package in R~\cite{chen2016xgboost}, which is an efficient implementation of gradient tree boosting. The measures were also compared to a baseline model, which always predicts the most frequent class in the training data (and provides a lower bound for prediction accuracy). The evaluation metrics were the accuracy, the ROC curve, and the area under the curve (AUC). To calculate these measures, we used 10-fold cross-validation. First, we randomly assigned every observation in the data set to one out of $k=10$ folds. For each fold ID from 1 to 10, we created a test set comprising all observations with matching fold ID and used all other observations for a training set. We then used the train set to learn model parameters, and we used the test set to evaluate prediction accuracy. The cross-validation pseudo-algorithm is available in the supplementary material, and the R code and datasets are available on \cite{chaves2020github}.

\subsection{Results}

Our evaluation dataset started with a total of 4,806 observations (178 participants, 27 evaluations per participant). From this total, participants skipped the question without answering for 11 observations. In 77 others, the participants signaled that they did not have a preference (the ``I don't know'' option). These observations were discarded from the analysis, resulting in a dataset with 4,718 observations. Each question-answer pair was evaluated from 24 to 35 times per construct. As we expected, participants overall preferred the answers from the original corpus, although the modified version was preferred for a few answers. A table with the number of votes per question is presented in the supplementary materials.

Figure~\ref{fig:acc-auc} shows the prediction accuracy and AUC plots for the four fitted models. Since participants generally preferred the original $FLG$ corpus answers, the prediction threshold is close to always predicting the most frequent class (original), which is particularly true for the random forest model. The prediction accuracy of \texttt{glmnet} and \texttt{xgboost} are only slightly better than the baseline. Nevertheless, the AUC plot indicates that the models are learning something important (the ROC curve plots are available in the supplementary materials), as the AUC values are consistently better than the baseline. Additionally, the \texttt{glmnet} model consistently selects the same variables to the k-folds.

\begin{figure}
\subfigure[Accuracy percentage]{\includegraphics[width=2.9in]{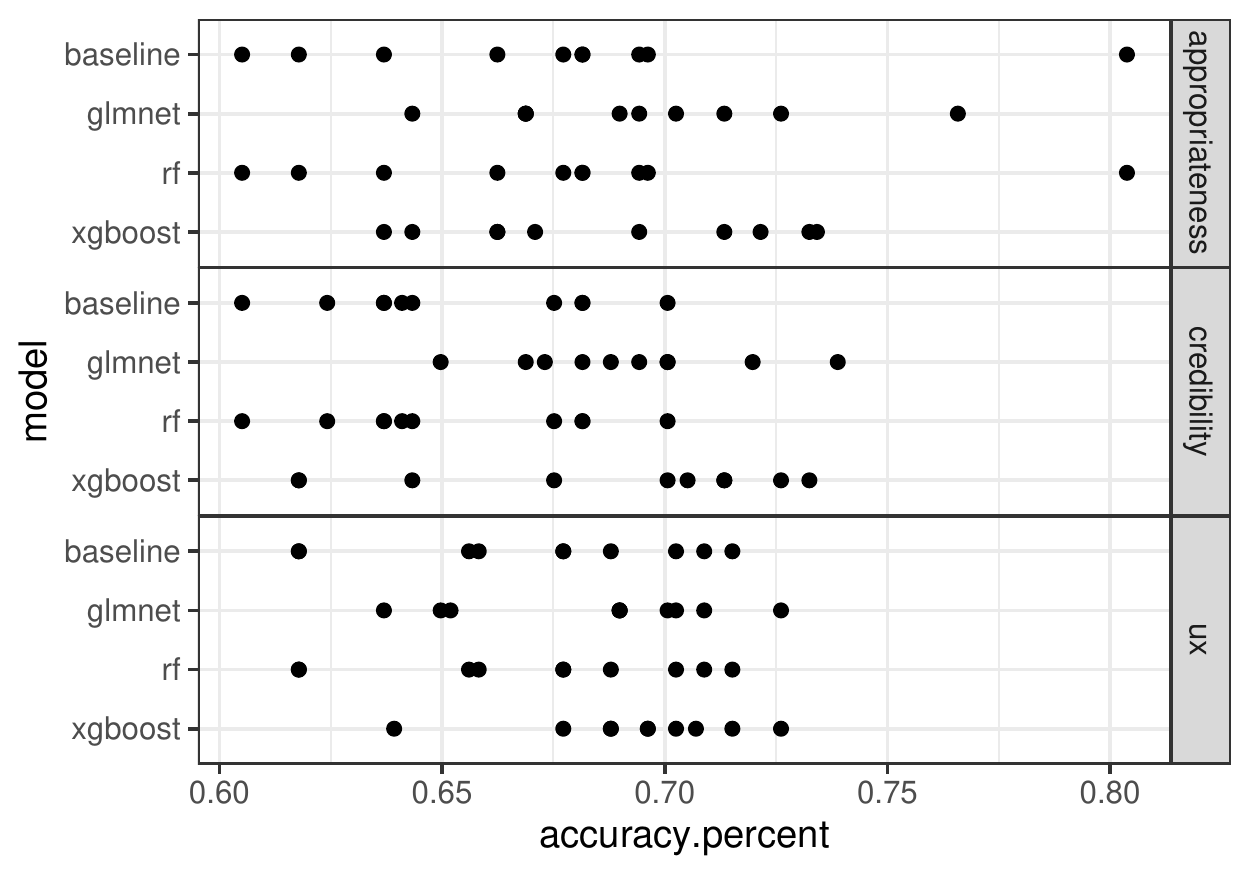}} \subfigure[AUC]{\includegraphics[width=2.9in]{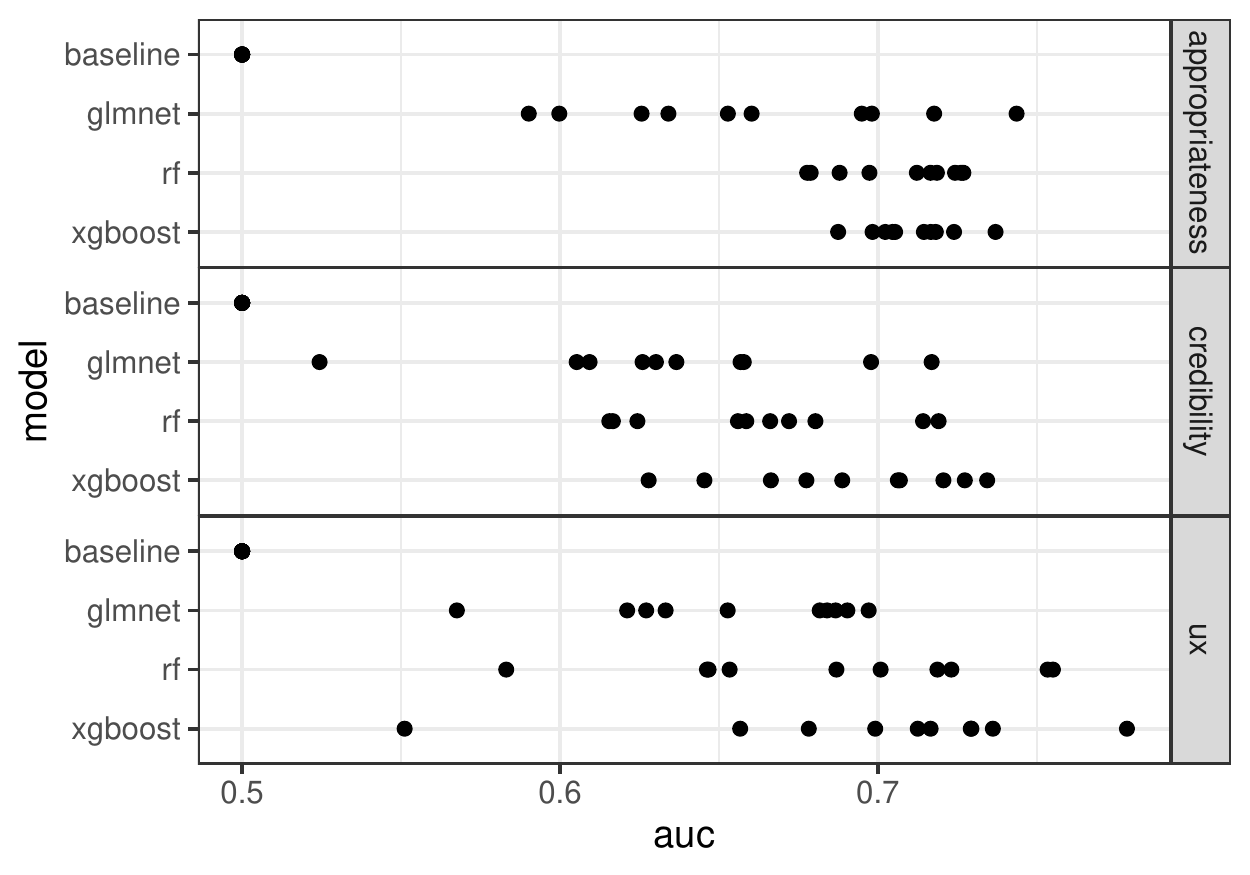}}
\caption{Accuracy (a) and AUC (b) results per model for each construct (appropriateness, credibility, and user experience). The baseline represents a model that always predicts the most frequent class (original). Accuracy percentage shows that glmnet, random forest, and xgboost perform only slightly better than the baseline model. AUC, however, is reasonably better than the baseline for the three models.}
\label{fig:acc-auc}
  \Description{(a) A plot with the accuracy percentage for every construct, comparing the four models (baseline, glmnet, rf, and xboost). (a) A plot with the AUC values for every construct, comparing the four models (baseline, glmnet, rf, and xboost).}
\end{figure}

Note that the non-linear models are not considerably more accurate than the linear model. Although there seems to be some non-linear trend in the data, it does not justify the complexity of the non-linear models. Hence, we used the results from \texttt{glmnet} model to interpret the coefficients and identify the linguistic features that determine user preferences.

\subsubsection{Coefficients}
\label{sec:coefficients}

Table~\ref{tab:coefficients} presents the coefficients of the linguistic features selected in six or more folds. The first and second columns indicate, respectively, the linguistic feature of interest and the sign of $\text{original}-\text{modified}$ calculation, which indicates whether one particular feature was increased or decreased in the text modification process. A positive sign (+) for a feature $f_i$ indicates that $\text{count}_{\text{original}}(f_i) > \text{count}_{\text{modified}}(f_i)$, while a negative sign (-) indicates that $\text{count}_{\text{original}}(f_i) < \text{count}_{\text{modified}}(f_i)$. The following three columns present the mean of the coefficients and the standard deviation for each construct. Features with negative coefficients increase the likelihood of the model predicting the original class. In contrast, features with positive coefficients increase the likelihood of the model predicting the modified class. The supplementary materials include plots of coefficients for each construct.

\begin{table}[ht]
    \centering
    \caption{Coefficients and standard deviation of the non-zero variables per construct. Only linguistic features were selected as relevant variables for predicting users choices, and most of the selected linguistic features are relevant for all the three constructs. The columns $\text{original}-\text{modified}$ represents whether the original answers have more $(+)$ or less $(-)$ occurrences of that particular feature. The dots indicate that the corresponding feature was not selected for that particular construct.}
    \label{tab:coefficients}
    \begin{tabular}{lcrrr}
        \toprule
        & & \multicolumn{3}{c}{\textbf{Mean of coefficients $\pm$ Std. Deviation}} \\
        Linguistic features & $\text{orig.}-\text{mod.}$ & \multicolumn{1}{c}{Appropriateness} & \multicolumn{1}{c}{Credibility} & \multicolumn{1}{c}{User Experience} \\ 
        \midrule
        Dim 1: Coordinating conj. clause           &  (+)  & -0.032 $\pm$ 0.003 & -0.031 $\pm$ 0.005 &	-0.038 $\pm$ 0.005 \\
        Dim 1: Attributive adjective               &  (+)  & -0.006 $\pm$ 0.001 & -0.006 $\pm$ 0.001 &	-0.007 $\pm$ 0.002 \\
        Dim 4: Conditional subordination           &  (+)  & -0.002 $\pm$ 0.002 &  0.005 $\pm$ 0.001 &   $\cdot$           \\
        Dim 1: Nouns                               &  (+)  &  0.002 $\pm$ 0.001 &  0.004 $\pm$ 0.001 &	 0.003 $\pm$ 0.001 \\
        Dim 1: Prepositions                        &  (+)  &  0.002 $\pm$ 0.001 &  0.001 $\pm$ 0.001 &	-0.001 $\pm$ 0.001 \\
        Dim 3: WH-relative clause (subj. pos.)     &  (+)  &  0.024 $\pm$ 0.004 &  0.034 $\pm$ 0.004 &   0.024 $\pm$ 0.004 \\
        Dim 1: Final preposition                   &  (+)  &  0.035 $\pm$ 0.007 &  0.038 $\pm$ 0.010 &	 0.017 $\pm$ 0.012 \\
        Dim 4: Suasive verbs                       &  (+)  &  $\cdot$           &  $\cdot$           & -0.015 $\pm$ 0.003 \\
        Dim 1: Causative subord.	               &  (+)  &  $\cdot$           &  $\cdot$           & -0.007 $\pm$ 0.005 \\
        Dim 2: Third person pronoun                &  (+)  &  $\cdot$           &  $\cdot$           &  0.012 $\pm$ 0.002 \\
        \midrule
        Dim 5: Adverbial--conjuncts                &  (-)  & -0.029 $\pm$ 0.007 & -0.017 $\pm$ 0.004 &	-0.056 $\pm$ 0.003  \\
        Dim 4: Prediction modals                   &  (-)  & -0.008 $\pm$ 0.001 & -0.006 $\pm$ 0.002 &	-0.011 $\pm$ 0.001  \\
        Dim 1: Contractions                        &  (-)  & -0.002 $\pm$ 0.001 & -0.008 $\pm$ 0.001 &	 $\cdot$            \\
        Dim 1: Second-person pronoun               &  (-)  & -0.002 $\pm$ 0.001 & -0.002 $\pm$ 0.001 &	 $\cdot$            \\
        Dim 1: First-person pronoun                &  (-)  &  0.007 $\pm$ 0.001 &  0.006 $\pm$ 0.002 &	 0.009	$\pm$ 0.004 \\
        Dim 1: Private verbs                       &  (-)  &  0.006 $\pm$ 0.001 &  $\cdot$	          &	 $\cdot$            \\
        Dim 1: Present verbs                       &  (-)  &  0.003 $\pm$ 0.001 &  0.005 $\pm$ 0.001 &   0.003 $\pm$ 0.001  \\
        Dim 1: That-deletion                       &  (-)  &  0.008 $\pm$ 0.003 &  $\cdot$           &   0.004 $\pm$ 0.003  \\
        Dim 3: Time adverbials                     &  (-)  &  0.018 $\pm$ 0.005 &  0.005 $\pm$ 0.003 &	 0.007 $\pm$ 0.005  \\
        Dim 1: Indefinite pronoun			       &  (-)  &  $\cdot$	        &  0.006 $\pm$ 0.004 &	 $\cdot$            \\
        \bottomrule
    \end{tabular}
\end{table}

Original answers have significantly more \textit{coordinating conjunction} clauses and \textit{attributive adjectives} than the modified versions. These features have a negative coefficient for all the three constructs, which indicates that frequent occurrences of these features increase the likelihood of original answers being chosen; participants are more likely to prefer answers in which these features are more frequent. The same conclusion applies to \textit{suasive verbs} and \textit{causative subordination}, although these features shows up as relevant only for the user experience construct.

Original answers also have significantly more \textit{nouns}, \textit{WH-relative clause (subject position)}, and \textit{final preposition}. These features have a positive coefficient for all three constructs, which indicates that frequent occurrences of these features increase the likelihood of modified answers being chosen. This outcome suggests that participants are more likely to prefer answers in which these features are less frequent. The same conclusion applies to \textit{third person pronouns}, but this feature is relevant only for the user experience construct.

Two features showed inconsistent outcomes across constructs. \textit{Conditional subordination} has a negative coefficient for appropriateness, but a positive coefficient for credibility. This outcome suggests that frequent occurrences of these features influence appropriateness positively while influencing credibility negatively. Additionally, \textit{prepositions} have a positive coefficient for both appropriateness and credibility, indicating that participants are more likely to choose the answers where these features are less frequent. In contrast, \textit{prepositions} have a negative coefficient for user experience, indicating that participants tend to choose answers where this feature is more frequent for this construct. However, when we aggregate the estimate to the standard deviation, it sums up to zero, suggesting that this outcome may be noise.

Modifications have significantly more \textit{adverbial--conjuncts} and \textit{prediction modals} than the original answers. These features have a negative coefficient for all three constructs, which indicates that frequent occurrences of these features increase the likelihood of original answers being chosen. This outcome suggests that participants are more likely to prefer answers in which these features are less frequent. The same inference applies to \textit{contractions} and \textit{second-person pronouns}, although these features did not show up as a relevant feature for user experience.

Modifications also have a larger number of \textit{first-person pronouns}, \textit{present verbs}, and \textit{time adverbials}. These features have a positive coefficient for all three constructs, which indicates that increasing the occurrences of these features increased the likelihood of modified answers being chosen. This outcome highlights the participants' preferences for answers in which these features are more consistently present. The same conclusion applies to \textit{private verbs}, \textit{that-deletion}, and \textit{indefinite pronouns}. However, \textit{private verbs} help to predict appropriateness only; \textit{that-deletion} predicts appropriateness and user experience only; and \textit{indefinite pronoun} is relevant for credibility only.

In conclusion, our results clearly show that \textit{the use of register-specific language has a significant impact on user perceptions of conversational quality for tourist assistant chatbots}. There is an association between register-specific use of particular linguistic features and perceived quality; linguistic features are stronger predictors of appropriateness, credibility, and user experience than individual characteristics of interlocutors (either assistants or users). The variables representing the tourist assistants who produced original answers, as well as those representing the individual participants and their social orientation, were not selected as predictors of user preferences regarding chatbot language use, since these factors were not relevant.

\section{Discussion}
\label{sec:discussion}

The results of the study presented here show that users are sensitive to conversational register and, specifically, that register has a significant impact on user perceptions of conversational quality; a chatbot that adopts the wrong conversational register risks losing credibility and acceptance by users. In this section, we discuss our findings and their implications for the design of register-specific language engines for chatbots.

User perceptions of conversational quality are important in chatbot design because the central interactional goal of chatbots is to fluidly interact using natural language. Chatbots are often deployed to perform social roles traditionally associated with humans, particularly in contexts where there may be consequences for a human if they choose to act on the chatbot's information. This means that user perceptions of chatbot competence and credibility are crucial for a chatbot's success. Some previous studies have found that appropriate language style is not relevant for determining user satisfaction as long as the user can understand the chatbot's answer, advising only that the chatbot's language style should be ``mildly appropriate to the service the chatbot provides''~\cite{balaji2019assessing}. We suggest, however, that the narrow focus on traditional usability metrics, such as effectiveness and efficiency, fail to adequately capture the broader context of user experience, which goes beyond simply comprehending a chatbot's utterance to, more importantly, whether the user trusts and ultimately uses the chatbot's advice. Specifically, user experience also involves the ``user's perceptions and responses that result from the use of the system''~\cite{iso2018ergonomics}, including emotions, beliefs, preferences, and perceptions, among others. The results presented here suggest that these crucial user perceptions are specifically shaped by \textit{how} information is conveyed--as characterized by the conversational register--rather than by the explicit information content of conversational exchanges. In this sense, our results support earlier behavioral observations that found that language fit can impact the quality of the interactions as well as the users' perceptions and behaviors toward the chatbots~\citep{jakic2017impact}. Importantly, the work presented here refines these observations by adapting register theory to provide a sound theoretical framework for concretely characterizing the concept of ``conversational style'' and analytically exploring how variations in the patterns of language impact user perceptions of chatbot quality, including factors critical for the user experience with chatbots, like appropriateness and credibility. Using register analysis to characterize different situations and how they map to the most appropriate register profiles also provides a way forward in the design of the next generation of chatbots; one could imagine a chatbot language engine that, given a particular situational profile for planned conversations, could automatically configure its language to present information in the most appropriate register. As a start in this direction, in the following we summarize some of the key insights regarding the complex relationship between conversational register and user experience.

\subsection{Key findings}

In this section, we begin with  several broad observations drawn from our findings before moving on to discuss specific linguistic features that have the greatest impact on user perceptions.

\subsubsection{Chatbot language design should account for register} Register is an established theory in the sociolinguistics domain (see e.g., \cite{biber1988variation, argamon2019register, biber2012register, biber1999longman}), and has been shown as a reliable predictor of language variation across conversational contexts. We aimed to explore the applicability of these results to human-chatbot interactions, develop a strong rationale for accounting for register in chatbot design, and provide a concrete mechanism for implementating theory into design practice. Our results show that, in terms of perceived appropriateness, credibility, and overall user experience, register characteristics are more relevant than individual preferences or personal habits, i.e., none of the covariates representing individual biases (participants, their social orientation, and answers' authors) were identified as predictors of participants' choices (see Section~\ref{sec:coefficients}). Thus, register characteristics can be seen as primary drivers for these perceptions, and designers should certainly consider to ensure chatbot acceptance and success.

\subsubsection{Certain linguistic features are preferred when efficiency matters} Like other information search interactions, the interactions in our corpus are goal-oriented. Particularly for en-route tourist information searches (see Section~\ref{sec:related-work}), users are often pressed for time; efficiency in finding the target information is critical~\cite{google2016how, tussyadiah2020review, wang2016smartphone}. Efficiency is also a priority in other task-oriented domains where chatbots operate, such as customer services~\cite{brandtzaeg2018chatbots}. Our statistical modeling revealed several linguistic features as particularly important in supporting compact, efficient information sharing. First, the linguistic feature \textit{coordinating conjunction} had the largest coefficient in our analysis; it is common in conversations due to real-time production constraints. Complex sentences in a conversation are often a linear combination of short clauses with a simple grammatical structure~\cite{biber1999longman}, usually connected by \textit{coordinating conjunctions}. $FLG_{mod}$ mimics $DailyDialog$ form, which portrays language that is more elaborated and carefully planned to achieve educational goals. The elaborated complexity leads to varying grammatical structure levels, which can reduce efficiency when providing information. Similarly, participants also often preferred answers with more verbs and fewer \textit{nouns}, which is a pattern present in the modified versions of the answers ($FLG_{mod}$). This outcome indicates a preference for active language rather than descriptive~\cite{biber1988variation}. For example, when comparing the answers

\begin{quote}
$FLG$: There are \$2 off coupons \textit{inside the visitor center behind the desk}.\\
$FLG_{mod}$: The visitor center has \$2 off coupons you can get.
\end{quote}

a participant stated that the first one ($FLG$) ``gives more information, but it's unnecessary'' and ``takes more time'' [LabP2].

Participants were also more likely to choose answers with fewer \textit{WH-relative clauses} and \textit{adverbial--conjuncts}. \textit{WH-relatives} are ``often an extra-piece of information that might be of interest''~\cite{biber1999longman}, for example, regarding the answers

\begin{quote}
    $FLG$: There is the Discovery Map, \textbf{which} is more geared toward visitors [...]\\
    $FLG_{mod}$: The Discovery Map is more geared toward visitors
\end{quote}

[LabP8] stated that $FLG_{mod}$ version was ``more clear'' with ``less fluff to the sentence.'' \textit{Adverbial--conjuncts} are used to connect sentences in discourse~\cite{biber1999longman}. Some are more frequent in face-to-face conversations (e.g., so, then, anyway), while others are more common in written language (e.g., however, therefore, although)~\cite{biber1999longman}. As $FLG_{mod}$ mimics $DailyDialog$ which focuses on language learning, the most common \textit{adverbial--conjuncts} align with the ones that are common in written registers. These conjuncts increase the formality of the answers~\cite{biber1988variation}, and interactive chatbot users may see these words as unnecessary, filler words. In the lab sessions, participants stated that ``when there is a lot of information, some filler words can be left out'' [LabP1]. For example, when comparing the answers

\begin{quote}
    $FLG$: You cannot leave it in 15-minute parking for an extended period of time. On the Amtrak side of the building, there is a paid parking lot.\\
    $FLG_{mod}$: You cannot leave it in 15-minute parking for more than that. \textbf{However}, the Amtrak side of the building has a paid parking lot you could use.
\end{quote}

a participant mentioned that the reason for the preference is that ``they don't have the `however,' and lead directly to the next sentence'' [LabP8]. Additionally, these conjuncts imbue a style of formality that may create distance between the chatbot and the user.

Finally, the preference for \textit{that-deletion} shown by the analysis is likely influenced by its frequent co-occurrences with \textit{suasive} and \textit{private verbs}. However, \textit{that-deletion} ``has colloquial associations and it is therefore common in conversations''~\cite{long1993modifications}, since conversations favor the omission of unnecessary words to accommodate real-time production constraints. Hence, user preferences for \textit{that-deletion} in our data could be associated with the preference for efficiency in conversations.

\subsubsection{Certain linguistic features impact the perception of human-likeness} The literature on human-chatbot interactions has extensively explored the need for chatbots to be ``human-like.'' On the one hand, scholars grounded in media equation theory~\cite{nass1994computers, fogg2003computers} have shown that people prefer agents who reflect human social and conversational protocols, e.g., conform to gender stereotypes associated with tasks~\cite{forlizzi2007interface}; self-disclose and show reciprocity when recommending~\cite{lee2017enhancing}; demonstrate a positive mood~\cite{hayashi2016effect}, and so on. On the other hand, overly humanized agents can create inaccurate expectations in users~\cite{gnewuch2017towards} and result in the ``uncanny effect''~\cite{appel2020uncanny, ciechanowski2018shades}, which eventually leads to more frustration when the chatbot fails to live up to these increased expectations~\citep{gnewuch2017towards}. However, the idea of assigning a social role to a conversational agent does not necessarily imply deceiving people into thinking the software is human; a chatbot can be clearly identified as such, but still benefit from approximating its conversational register to the patterns of human-human communication. This study brought to light a crucial aspect of human-chatbot interaction, namely the need for balancing the chatbot's anthropomorphic clues--several specific findings in our analysis support this observation.

In the Natural Language Generation field, the aggregation of sentences using \textit{coordinating conjunctions} is commonly used to increase fluency and readability~\cite{reiter2000building}. According to~\citet{reiter2000building}, aggregation of sentences can be de-emphasized if the text is obviously produced by a computer; this suggests that participants would not care about slightly stilted language, since they were told that the answer was produced by a chatbot. Our analysis reveals, however, that users have a preference for language that is more human-like, with fewer pauses and more coordination.

\textit{First-person pronouns} can also increase human-likeness, although the plural form is preferable. The singular form (``I'') unambiguously indicates the speaker, whereas referring to the speaker's identity in the plural form (``we'') varies according to the context~\cite{biber1999longman}. Choosing between singular or plural forms is a strong indicator of the identity that the chatbot conveys. When the chatbot says ``I,'' it clearly refers to itself, but when it says ``we,'' it can be interpreted as a general reference to its social category. Using ``we'' softens the role of the chatbot and highlights its representative role of a broader entity (e.g., professional tourist assistants, visitor center's representatives, Flagstaff's tourism personnel). Both singular and plural forms convey personal involvement, but ``we'' may demonstrate more credibility because the chatbot is more likely to be recognized as part of a community of knowledgeable individuals. Although both singular and plural pronouns are measured under the same linguistic feature, our evidence shows that participants preferred the plural use of this feature. As often occurs in $DailyDialog$, the singular form (``I'') co-occurred with the \textit{prediction modal} ``would'' in $FLG_{mod}$ to make suggestions or to give advice. Noticeably, participants preferred answers where \textit{prediction modals} are less frequent, which indicates that the singular form of \textit{first-person pronouns} is unlikely to influence the users' preference for this feature. Quotes from lab sessions also support that participants preferred the plural form. For example, regarding the singular form, [LabP1] stated: ``I don't like when the chatbot says `I,' it seems the developer is trying to trick me to think the bot has opinions. When chatbots use `I' it sounds too much like pandering.'' In contrast, when comparing the answers 

\begin{quote}
    $FLG$:``There are 50 miles of trails within Flagstaff [...]''\\
    $FLG_{mod}$ ``We have 50 miles of trails within Flagstaff [...]''
\end{quote}

[LabP3] stated that ``the use of the word `we' makes it more personable than simply saying `there is this' '' [LabP3].

Since the language in the baseline corpus ($FLG$) was human-written and not tailored to represent the identity of a chatbot, participants likely perceived some of the chatbot's answers as overly human-like. As these findings reveal, chatbot language should conform to the expectations of its social category, as previous literature has suggested~\cite{gnewuch2017towards}. Still, the register of chatbot conversation must also consider its artificial nature, particularly positioning the agent as a representative of a broader entity. As register theory suggests~\citep{conrad2009register}, the interlocutors' identities and the relationship among them are influential parameters when defining the interactional context. Therefore, tailoring the chatbot's language to the appropriate register includes not only adapting to the language of the professional category it represents, but also revealing the chatbot's social identity as an artificial agent. These observations strengthen the relevance of conversational register as a theoretical foundation for the design of chatbot utterances.

\subsubsection{Certain linguistic features impact the perceived level of personalization} Previous literature has shown the benefits of personalized interactions with chatbots~\cite{thies2017how, shum2018eliza, duijvelshoff2017use}, particularly in domains where the chatbot needs to build rapport and trustful relationships with the user, such as in financial services~\cite{duijst2017can}, companion (or buddy) chatbots~\cite{thies2017how, shum2018eliza}, and recommendation systems~\cite{cerezo2019building}. At the same time, users might feel uncomfortable with some aspects of personalized content~\cite{thies2017how}. In our study, tourist assistants did not provide personally tailored information, as they had few or no clues about the tourists' identities or preferences due to the text-based nature of the conversations. In this inherently rather impersonal content, our results showed that participants preferred general rather than personalized information.

Participants preferred lower frequencies of occurrences of \textit{second-person pronouns} (e.g. ``you''), which are used as a direct address to the user~\cite{biber1999longman}. Giving that information search interactions focus on the assistant providing information without necessarily sharing a personal relationship with the user, it may sound inappropriate for a chatbot to use \textit{second-person pronouns}, rather than simply stating the information. In the lab sessions, for instance, a participant stated that ``[the chatbot] saying `you' implies a lot of personalization but in a pressuring type of way'' [LabP6]. \textit{Second-person pronouns} (as well as singular \textit{first-person pronouns}) often co-occur with \textit{contractions}, which may justify the participants preferences for answers with low frequency of \textit{contractions}.

The appropriateness of \textit{conditional subordination} also relates to personalization. The use of conditional clauses as a mechanism for inserting suggestions, requests, and offers is common in conversation~\cite{biber1999longman}. The tourist assistants in $FLG$ used \textit{conditional subordination} to offer different options to the users (e.g., \textit{``if you..., you can/will/would'')}, since their individual preferences were unknown. In the lab session, [LabP1] mentioned ``prefer the [original answer] because it says `if you stop' rather than 'I'd suggest you stop'.'' Moreover, \textit{conditional subordination} helps to frame the subsequent discourse~\cite{biber1999longman}, which was also observed by [LabP1]: ``I like the 'if you are looking for maps' as it sets up the scenario that this information would be useful for'' [LabP1]. However, the \textit{conditional subordination} is negatively associated with credibility, which may indicate that when the chatbot gives options, it sounds as if it is not confident about the information provided. It is important to observe the impact of the varying communication purposes here. Although the chatbot is presented as an information provider, the tourist assistants interpreted some tourist questions as requests for recommendations (see more in Chaves et al. ~\cite{chaves2019identifying}) rather than information. Recommendations are more inherently personalized than information search results, and consequently require more personalization in their expression (see e.g.,~\cite{gavalas2011web, ricci2011introduction, komiak2006effects}). Thus, participants may have expected that the tourist assistants would provide more personal, tailored content rather than conditional options. Clearly, the dynamics that shape these interactions are subtle, and the influence of sub-registers, i.e., the variation in language use to match specific communicative purposes, will need to be explored in more detail to evaluate these assumptions.  

\subsubsection{Avenues for future investigation} the study we presented in this paper accomplished its primary goal and, at the same time, exposed deeper complexities that point to the need for further exploration. For instance, we found inconsistent results regarding \textit{private verbs} when comparing the cross-validation results to the quotes from lab sessions' participants. The cross-validation model found an association between the high frequency of \textit{private verbs} and appropriateness, with no effect on the other two constructs (credibility and overall user experience). However, several quotes from lab session participants suggest that \textit{private verbs} did make certain answers less credible. For example:

\begin{quote}
`` `I guess' is not the tone I want.'' [Lab1P]

``I don't like the bot saying `I guess', it sounds passive aggressive'' [LabP2]

``I don't like how the [modified answer] says `okay, I believe'. This makes it sound like it doesn't know.'' [LabP4]
\end{quote}

Considering such qualitative feedback from these participants, we believe that the lack of any negative influence of \textit{private verbs} on credibility shown by the analysis may be conditioned by their co-occurrences with other features; this needs to be further investigated.

Our results also showed a positive association between \textit{attributive adjectives} and all the three evaluated metrics (appropriateness, credibility, and user experience). However, this association may be too coarse-grained, as the way in which \textit{attributive adjectives} were often used in the specific conversational context of tourism advising is somewhat atypical. The typical use of \textit{attributive adjectives} in conversation is to describe some physical attribute of an object~\cite{biber1999longman}, e.g., ``new,'' ``big,'' or ``smelly.''  In contrast, \textit{attributive adjectives} in the $FLG$ corpus were more often used \textit{to classify} rather than describe the corresponding \textit{nouns}. For example, common \textit{attributive adjectives} are ``local business,'' ``national park,'' and ``natural landmark.'' Using classifying \textit{attributive adjectives} adds detail to the information without loss in efficiency. Participants mentioned that the \textit{attributive adjective} ``makes [the answer] more interesting'' [LabP7], explaining the association with quality metrics, but this may apply only to classifying \textit{attribute adjectives}. Here too, further investigation will be needed to clarify the validity of the observed association.

Finally, \textit{suasive verbs} are typically used to express the degree of certainty associated with the information that the sentence communicates~\cite{biber1999longman}. For example, when the tourist assistant says ``I recommend,'' it represents how much it believes the tourist should take that advice. [LabP3] observed this fact by stating that `` `I would recommend' seems like more of a suggestion.'' Our analysis showed that \textit{suasive verbs} influence the overall user experience, but were not shown to make the language more credible or appropriate. Closer analysis shows that this association, too, deserves further investigation. In our data, \textit{suasive verbs} co-occurred mainly with the singular \textit{first-person pronoun}. As noted earlier, the use of plural \textit{first-person pronouns} was clearly linked to credibility; further investigation should evaluate whether the use of plural \textit{first-person pronouns} with \textit{suasive verbs} would increase their impact on credibility.

\subsubsection{Applicability of results to other domains} The register theory aims to link the appearance of certain linguistic features in utterances to the situational parameters of the conversation~\cite{conrad2009register}. Although characterizations of situational parameters and their detailed impacts on the selection of conversational register may continue to evolve and be refined over time, the patterns of language should be similar in domains that share similar situational parameters. For instance, information search, the core interactional purpose of the interactions studied in this research, is also a common interactional purpose in customer services interactions, i.e., two participants working to fulfill an information request. The two domains share other situational parameters as well, such as channel, production, and setting. ~\citet{abu2004evaluation} observed that conversations from a corpus of Spoken Professional American English portray more \textit{coordinating conjunctions}, \textit{subordinating conjunctions}, and \textit{plural personal pronouns} than transcripts from the chatbot ALICE. These same linguistic features were selected in our analysis as predictors of appropriateness, credibility, and user experience (see Section~\ref{sec:coefficients}). In contrast, we expect that a sales chatbot would require more persuasion (features in Dimension 4, see supplementary materials for the full list of linguistic features), and recommendation-based chatbots would require more personalization, as we discussed previously in this section. In any case, a register analysis, as presented in this paper, could be used as a tool to analyze the conversation register used by expert humans in such conversational scenarios, and to identify specific linguistic features of that register that are relevant to producing the desired impact on user perceptions.

\subsection{Implications for chatbot design}

This research has important implications for designers of the next generation of chatbots. For chatbots that find and retrieve knowledge snippets from external sources, utterances should be adapted to the conversational situation in which the chatbot is embedded. This is not generally done in the current generation of chatbots; it is not uncommon to find chatbots that extract and present information directly from websites, books, or manuals without any adaptation. For example, Golem\footnote{Available in Facebook Messenger at http://m.me/praguevisitor. Last accessed: June, 2020} is a chatbot designed to guide tourists through Prague (Czech Republic); its utterances are extracted from an online travel magazine\footnote{https://www.praguevisitor.eu} without any adaptation to the new interactional situation (which differs in production, channel, and setting). Moreover, new generations of chatbots will be expected to dynamically generate their own custom-constructed utterances. They will need sophisticated language engines that are able to dynamically adapt their conversational register to changing situational parameters. In this context, corpora such as $DailyDialog$ are likely to become a baseline for training the conversation models~\cite{galitsky2019chatbot} at heart of such language engines; our study emphasizes that designers should carefully ensure that the register found in any corpus used to train such models matches the optimal register implied by the situational parameters, or that the learning algorithms can adapt the language accordingly.

This paper provides a list of linguistic features that conform with user expectations about chatbots' language use in the context of tourism information searches, which can be directly applied to the design of chatbots for this domain. Researchers could leverage these outcomes to evaluate the application of these results to similar interactional situations in other domains. Section~\ref{sec:discussion} discusses the relation of these outcomes with other task-oriented domains, such as information searches and customer services. Additionally, the methodology presented in this paper can be applied to other domains that are more distant in terms of interactional situations from $FLG$, aiming to identify the associations between new interactional situations and core linguistic features used in the domain. As our study shows, the analysis does not necessarily require a large number of individuals to identify the linguistic features that characterize the register of a particular situation or domain.

Finally, the parallel corpora $FLG$ and $FLG_{mod}$ are available for researchers and practitioners who are interested in (i) developing chatbots for tourism information searches, as they comprise a set of frequently asked tourism questions with register-specific answers; or (ii) research on natural language generation that requires parallel data. The corpora and associated materials are available online~\cite{chaves2020github}.

\section{Limitations}
\label{sec:limitations}

Register characterization relies on the multidimensional approach proposed by \citet{biber1988variation, biber1995dimensions}, which is the main theoretically-motivated approach taken within register analysis~\cite{argamon2019register}. Other approaches, such as register classification~\cite{argamon2019register}, could be explored in the context of human-chatbot interactions. Additionally, the register analysis also relies on Biber's grammatical tagger to automatically tag the linguistic features. The tagger has been used for many previous large-scale corpus investigations, including multidimensional studies of register variation (e.g., ~\cite{biber1995dimensions, biber1988variation, conrad2014multi}), The Longman Grammar of Spoken and Written English~\cite{biber1999longman}, and a study of register variation on the Internet~\cite{biber2016using, biber2017register}. Although this tagger achieves accuracy levels comparable to other existing taggers~\cite{biber1988variation}, mis-taggings are possible. To mitigate this effect, we manually inspected a small subset of tagged files to search for mis-tags that could potentially impact the outcomes.

We performed manual linguistic modifications to produce the $FLG_{mod}$, which inherently introduced a subjective element in the exact choice of changes applied to shift the register. We mitigated this threat by manually inspecting $DailyDialog$ corpora for every feature we modified, to understand the function of the feature in the corpus, and produce modifications using similar patterns. We also performed a validation with human participants for content preservation and quality of modifications.

We included in the cross-validation model only features that vary between $FLG$ and $DailyDialog$, and therefore were manipulated during the text modification. We considered that the linguistic features that do not significantly vary across the corpora are the standard in those particular contexts and are unlikely to signal users' preferences. We claim that $DailyDialog$ is an appropriate dataset to be used in this study since it has been widely used in natural language and dialogue generation research (see, e.g., \cite{zhao2018unsupervised, shen2018improving, gu2018dialogwae}), and might eventually become a baseline for learning conversation models~\cite{galitsky2019chatbot}. However, it is also important to compare $FLG$ against corpora produced in other interactional situations to evaluate varying sets of features and confirm the inferences presented in this paper.

The register analysis presented in this paper was based on counts of the occurrences of features, normalized per 10,000 words. However, it does not consider sentence structure, i.e., where the features occur in the sentence. Additionally, because linguistic features are best understood in terms of co-occurrence patterns, it is important to extend this study to consider the effect of the linguistic features individually and the effect of features that typically co-occur with them. Future research is needed to expand the analysis to incorporate these aspects.

This research is performed in the context of conversations in American English. The core linguistic features and their usage change from one language to the other; thus, these results may not apply to other languages and further investigation is necessary.

Three tourist assistants, all female, answered tourist questions in the $FLG$ corpus, and tourists were recruited in Flagstaff, AZ, USA. To increase the diversity of tourists' questions, we mined questions from websites, as discussed in Section~\ref{sec:data-collection}. Concerning tourist assistants, an interesting extension of this study would include hiring tourist assistants with a more diverse profile and including questions about other touristic cities. It is important to note that, even with only three tourist assistants, we were able to identify the impact of certain linguistic features on the three metrics (appropriateness, credibility, and overall experience) used to represent perceived conversational quality; this suggests that a very large corpus is not necessary to identify the core linguistic features of chatbot dialogues that influence user perceptions.

Finally, with limited qualitative observations from our in-lab sessions to support the quantitative findings, our ability to draw conclusions based on participants' statements is incomplete. The purpose of adding this qualitative element was to augment and clarify our quantitative findings, by identifying participants' impressions aligned with our quantitative analysis and comparing the impressions of our participants to the interpretations of linguistic features we find in the register literature, such as in the Longman Grammar~\cite{biber1999longman}. A deeper qualitative investigation is needed to draw stronger conclusions about motivations behind participants' choices.

\section{Conclusions}
\label{sec:conclusions}

This paper focuses on investigating the impact of chatbot language use on user perceptions of language appropriateness, credibility, and the overall user experience. We collected two corpora of conversations between tourists and tourist assistants in different interactional situations and compared the language variation in these corpora by adapting techniques associated with register analysis, which are well-established by sociolinguists. Based on this analysis, we produced two parallel corpora of conversational exchanges that were equivalent in informational content but differed in linguistic form. We then used the parallel corpora to perform a study of the impact of register on user preferences, asking participants to rate parallel responses draw from the corpora on three metrics of perceived conversational quality: appropriateness, credibility, and overall experience. Participants were told that a tourist assistant chatbot had generated all responses.

Our results revealed statistically relevant associations between certain linguistic features present in utterances within the two corpora and user perceptions of appropriateness, credibility, and overall user experience. Additionally, the results also showed that the linguistic features are a stronger predictor of this association than the variables representing individual biases (participants, their social orientation, and answers' authors). This outcome strongly suggests that attention to an appropriate conversational register is an important factor for the perceived quality of chatbot conversations and, therefore, critical to future chatbots' success. Although our study focused on the tourism domain, we expect these outcomes to be applicable to other interactions that share similar situational parameters (e.g., different information search scenarios). More generally, this study demonstrates that the theoretical foundation of register analysis introduced in this paper can be an effective tool for characterizing the conversational register used in other target domains, and can systematically expose the specific linguistic features within conversational utterances that most strongly impact user perceptions of conversational quality. This makes register a promising cornerstone for rationalizing the design of future generations of chatbots, by reproducing this study for other domains, as well as developing empirical studies that extend this study to other aspects of register theory, such as the impact of sentence structure. Specific future directions for our research will focus on evaluating the effect of register-specific language within dynamically-generated interactions with chatbots on user perceptions of conversational quality.

\begin{acks}
To Caitlin Abuel and Tyler Conger, NAU CS undergraduate students, for the contributions to the recruitment and qualitative data collection during the lab sessions. This work is supported by the National Science Foundation under Grant No.: 1815503.
\end{acks}

\bibliographystyle{ACM-Reference-Format}
\bibliography{base}


\end{document}